\documentclass [pra,noeprint,reprint,groupedaddress,showpacs]{revtex4-1}
\usepackage{graphicx}
\usepackage{dcolumn}
\usepackage{amsmath}
\usepackage{amsfonts}
\usepackage{amssymb}
\usepackage{bm}
\usepackage{nicefrac}
\usepackage{color}
\usepackage{float}
\usepackage{sidecap}
\usepackage{mathtools}
\usepackage{cancel}
\graphicspath{{figures/}}
\usepackage{hyperref}
\usepackage[none]{hyphenat}
\hypersetup{backref,
	pdfstartview=Fit,
	colorlinks=true,
	citecolor=blue,
	urlcolor = blue,
	linkcolor = blue}
\begin{document}
	
\title{Driven quadrature and spin squeezing in a cavity-coupled ensemble of two-level states}

\author{Haitham A.R. El-Ella}
\email{haitham.el@gmail.com}

\affiliation{Department of Physics, Lund University, SE-22100 Lund, Sweden}

\begin{abstract}
The generated magnitude of quadrature squeezing in a cavity-coupled ensemble, which is continuously driven using a coherent off-axis field, is theoretically explored. Using a truncated set of equations of motion derived from a Dicke Hamiltonian, steady-state quadrature squeezing of the cavity field is numerically calculated to approach a limit of -3 dB, while frequency-modulated quadrature squeezing approaches a limit of -14 dB, in the absence of pure dephasing, and as a function of the ensemble's size and detuning. The impact of pure dephasing on steady-state quadrature squeezing is shown to be mitigated by increased detuning of the driving field, while frequency-modulated squeezing is only shielded in a regime where the cumulative coupling and driving rates are in excess of the pure dephasing rate. Spin-squeezed entanglement is also calculated to occur simultaneously with weakly driven frequency-modulated quadrature squeezing.
\end{abstract}
\maketitle

\section{Introduction}
Quadrature squeezed light is an important experimental resource in quantum optics, with a number of applications ranging from enhancing interferometry beyond the shot-noise limit \cite{Schnabel2017, Aasi2013}, to its use in generating entangled continuous-variable states for quantum information protocols \cite{Asavanant2019,Larsen2019}. Due to its utility, there is justifiable motivation to not only generate larger squeezing magnitudes, necessary in particular for fault-tolerant continuous-variable quantum computing \cite{Walshe2019}, but also in expanding its bandwidth \cite{Singh2019}, and in optimizing its experimental efficiency and integrability \cite{Arnbak2019}. By making squeezed light sources more accessible and practically implementable, their benefits may be reaped in both routine spectroscopy and interferometry \cite{Michael2019, DeAndrade2020}, while further spurring the development of hybrid continuous and discrete variable quantum information protocols \cite{Andersen2015}, and optical sensing schemes that go beyond the classical limits \cite{Degen2017}. 

State-of-the-art sources of quadrature squeezed light are based on cavity-assisted $\chi^{(2)}$ parametric down-conversion \cite{Vahlbruch2016}, while much effort is currently being invested in developing alternative on-chip integrated sources based on $\chi^{(3)}$ four-wave mixing schemes \cite{Hoff2015,Vaidya2020}. An alternative quadrature squeezing mechanism, which is technically simpler but less explored, is based on the resonant fluorescence of weakly driven optical dipoles, first proposed by Walls and Zoller \cite{Walls1981}. 

The maximum measurable degree of quadrature squeezing in free space from such two-level systems is predicted to be in the order of -1.25 dB, without accounting for optical losses and realistic detection efficiencies. Experimental attempts so far have successfully substantiated this prediction; however, due to limited detection efficiencies and cumulative optical losses, the measured quadrature squeezing has been far below the predicted value, ranging between the orders of -10 to -100 mdB for cavity-coupled atoms and quantum dots, respectively \cite{Ourjoumtsev2011, Schulte2015}. 

Compared to the measured -15 dB from state-of-the-art parametric cavity systems \cite{Vahlbruch2016}, and considering the -15 to -17 dB desired for fault-tolerant continuous variable quantum computing \cite{Walshe2019}, the motivation for pursuing resonance-fluorescence based quadrature squeezing lies rather in the possible technical advantages and accessible wavelengths. The appeal of their potentially small technical footprint, and in providing squeezed light sources at wavelengths towards the higher energy end of the visible spectrum, makes exploring this approach worthwhile. The latter point is particularly interesting, given the technical challenges of frequency converting squeezed vacuum states \cite{Vollmer2014}, and the difficulty of engineering suitable nonlinear systems for generating such states at wavelengths shorter than 600 nm. 

Quadrature squeezing through resonance fluorescence is based on an established proportionality between the scattered field's quadrature fluctuations and the dipole moment's fluctuations, such that they may be considered interchangeable \cite{Meystre1982, Collett1984, Wodkiewicz1987}. When considering an ensemble of noninteracting dipoles, this relationship may be transposed into a relationship between the collective angular momentum operator and the far-field quadrature, thereby highlighting a possible link between far-field quadrature squeezing and ensemble spin squeezing \cite{Kitagawa1993}. In turn, given the direct relationship between spin-squeezing and multipartite entanglement \cite{Sorensen2001, Haakh2015}, any observable nonclassical fluctuation of the far-field quadratures can be considered an unambiguous witness of multipartite entanglement, for certain experimental configurations.

Spin-squeezed states are an important class of metrological probes, usually employed in interferometric schemes that revolve around assessing a phase shift of the collective spin state, imparted by an external physical quantity. The smallest uncertainty when measuring such phase shifts, and therefore the spin state's ultimate sensitivity, is directly proportional to its degree of multipartite entanglement \cite{Toth2009,Vitagliano2018}. Experimentally determining large-scale entanglement is a principal goal in many areas of quantum information science. Therefore, exposing and delineating relationships between metrologically useful multipartite entangled states and directly measurable quantities, such as the quadrature fluctuations of coupled fields in this work, is fundamentally informative. 

Here, an indirectly driven cavity-ensemble system is numerically explored to determine the conditions needed for generating quadrature-squeezing magnitudes beyond the free-space limit, and the consequential degree and type of spin squeezing. Analogous to the cavity-mediated detuned scheme studied in \cite{Grunwald2013} and the Raman-based scheme in \cite{Dimer2007}, the cavity-ensemble system is numerically solved for varying detuned configurations that address the side-bands, or dressed states, of the coupled system, with the various rates and detuning framed in relation to the two-level state's longitudinal relaxation rate.  

The numerical results highlight how the generation of both steady-state and frequency-modulated quadrature squeezing of a cavity field can exceed the free-space limit of single two-level systems while accounting for pure-dephasing. Furthermore, they highlight how the generation of frequency-modulated quadrature squeezing can simultaneously generate entangled spin-squeezing. 

The investigation begins with delineating the studied Hamiltonian, the assumptions made, and the considered solid-state ensemble systems in Sec.\ref{Sec2}. This is followed by a discussion of the numerical steady-state results for quadrature squeezing from off-axis-driven and cavity-coupled single, and ensemble, two-level emitters in Sec.\ref{Sec3}. Finally the numerical results for generated frequency-modulated quadrature squeezing and the simultaneous occurrence of entangled spin squeezing are presented and discussed in Sec.\ref{Sec4}. 

\section{Hamiltonian \& Fluctuations} \label{Sec2}

\subsection{System Hamiltonian and dynamics}

\begin{figure*}
	\includegraphics[scale = 0.9]{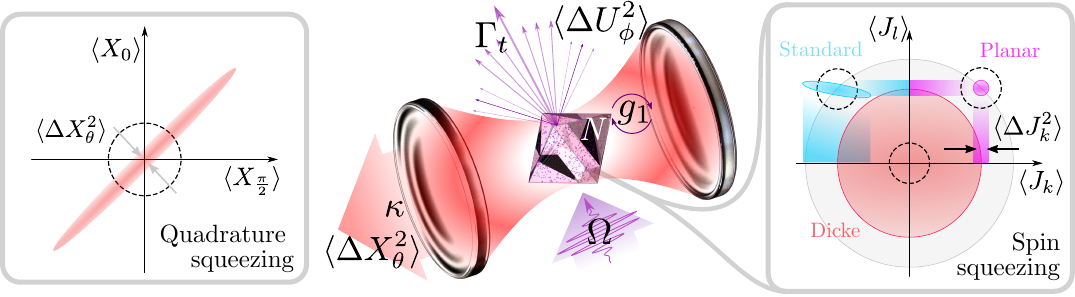}
	\centering
	\caption{A simplified schematic describing the considered system parameters. An  $N$-sized ensemble is placed within an optical cavity with a decay rate $\kappa$, and is collectively coupled with a rate $g_1\sqrt{N}$. The ensemble is driven coherently with an off-axis field at a rate $\Omega$, and decays with a total dephasing rate $\Gamma_t$. The system's intra-cavity and off-axis field's quadrature fluctuations $\langle \Delta X_\theta^2\rangle$ and $\langle \Delta U_\theta^2\rangle$ are squeezed along one direction in phase-space.  The ensemble spin-state fluctuations $\langle \Delta J_k^2 \rangle$ may be squeezed simultaneously along one or two of the orthogonal coordinates of the ensemble Bloch sphere, resulting in three possible types of squeezed spin states. The dashed circles represent the shot-noise or spin-noise uncertainty limit.}
	\label{fig1}
\end{figure*}

A two-level system, representing either an optical or magnetic dipole, is typically defined using pseudo-spin operators for a ground and excited state basis $\{\vert g\rangle, \vert e\rangle \}$ separated by an energy $\hbar\omega_0$, such that $\sigma = \vert g \rangle \langle e \vert$, $\sigma = (\sigma^\dagger)^\dagger$, and the commutation $[\sigma^\dagger,\sigma] = \sigma_z$. A Hermitian quadrature operator $U_\phi$ for a dipole may be similarly defined to that of a single optical mode $X_\phi$, considering a relative measurement phase or the optical field's instantaneous phase ($\theta$ or $\phi$):
\begin{align}
\label{1}
U_\phi = \big(e^{i\phi}\sigma + e^{{-}i\phi}\sigma^\dagger\big),~~X_{\theta} =\big(e^{-i\theta}a+e^{i\theta}a^{\dagger}\big).
\end{align}  
Assuming that an ensemble's constituents only interact indirectly via the external fields, the energy and dynamics of a cavity-coupled ensemble may be described by a Dicke Hamiltonian: 
\begin{align}
\label{2}
\mathcal{H}/\hbar &= \omega_c a^{\dagger}a + \frac{1}{2}\sum_k^N \omega_k\sigma_{kz} +\sum_k^N g_k X_{0} U_{\frac{\pi}{2},k},
\end{align}
where $\omega_c$ and $\omega_k$ are the cavity and dipole transition frequencies, and $g_k$ is the coupling strength of the dipole and cavity mode. In the scenario discussed here, the dipole ensemble is considered to be directly driven by an off-axis field which bypasses the cavity input, as described in \cite{Grunwald2013}. This is easily facilitated using a ring-based cavity configuration, but may also be implemented using a linear cavity via either off-axis excitation (requiring consideration of the relative dipole alignment with the external driving field and the cavity field), or detuning of the cavity from the dipole's and external field transition. 

Such a system is schematically pictured in Fig.\ref{fig1}. This configuration ensures that the cavity remains a passive element which acts as a coherence `purifier' of the ensemble's transition, in the sense of which the coherent cavity-coupling rate outcompetes the incoherent decay rates of the ensemble. For an appropriately engineered cavity system, this configuration can collect a large part of the ensembles emitted fluorescence while avoiding the bandwidth restrictions of driving the ensemble through the cavity itself, in addition to being experimentally convenient for distinguishing the driving field light from the cavity-transmitted light during detection. 

The driving-field of the ensemble is described using a dipole approximated semi-classical term:     
\begin{align}
\label{3}
\mathcal{H}_\Omega/\hbar &= \frac{\Omega}{2}\big(e^{i\omega_lt} + e^{-i\omega_l t} \big)\sum_k^N U_{0,k},
\end{align}
which is defined as a function of a driving Rabi frequency $\Omega$ and a field frequency $\omega_l$, and for which $\Omega$ represents a product of a linearly polarized plane-wave electric field and a linear transition-moment of the two-level state. 

The resulting system dynamics are calculated by numerically integrating the system's Markov approximated master equation $\dot{\rho}$, while accounting for a cavity decay rate $\kappa$, the radiative longitudinal relaxation rate $\gamma_1$, and pure-dephasing rate $\gamma_2^*$, as detailed in the Appendix. 

A thermal starting configuration of the system (relevant for e.g. magnetic dipoles) is consistently used here for all numerically derived results, such that $\{\langle\sigma_z(0)\rangle,\langle a(0)\rangle\} = \{0,0\}$, and $\langle a^\dagger a(0) \rangle = \bar{n}$. It should be noted that identical results are obtained when considering $\langle\sigma_z(0)\rangle = -1$, which would be the expected case for an ensemble of optical dipoles.     

When deriving the coupled equations of motion, the system is treated symmetrically as carried out in \cite{Meiser2009, Henschel2010}, such that for all ensemble constituents $k$, single dipole expectation values and their correlation with the cavity field are considered equal $\langle\sigma_k\rangle = \langle\sigma_1\rangle$, $\langle a^\dagger\sigma_k\rangle = \langle a^\dagger\sigma_1\rangle$, while all pairs of spins are designated as  $\langle \sigma_k^\dagger\sigma_j\rangle = \langle \sigma_1^\dagger\sigma_2\rangle$  for all $j \neq k$, adhering to the commutation relation $[\sigma^\dagger_j,\sigma_k] = \sigma_{jz}\delta_{jk}$. The generation of the equations of motion is an indefinite procedure which is truncated using a cumulant expansion approach \cite{Kubo1962, Meiser2009, Henschel2010}, as described in the appendix. 

A rotating frame with the driving field frequency $\omega_l$ is considered, along with the rotating wave approximation. However, this is not employed for the cavity coupling term, as the rapidly oscillating terms become non-negligible for increasing ensemble sizes, especially considering the possibility of the collective coupling strength approaching the transition frequency. 

\subsection{Quadrature squeezing}
The variance of an operator's fluctuations is defined with respect to its expectation value such that:
\begin{align}
\label{4}
\Delta a =& a-\langle a\rangle,\nonumber
\\
\big\langle\Delta a^2\big\rangle =& \langle a^2\rangle -\langle a\rangle^{2}.
\end{align}
For a single-mode field, its quadrature fluctuations is defined as:
\begin{align}
\label{5}
\langle\Delta X_{\theta}^{2}\rangle &= 2\Big(\langle\Delta a^\dagger\Delta a\rangle +\Re\Big\{e^{-i2\theta}\langle\Delta a^2\rangle\Big\}\Big) + 1.
\end{align} 
This expression consists of a coherent and incoherent contribution $\langle \Delta a^2\rangle$ and $\langle \Delta a^\dagger\Delta a\rangle $, respectively, which are both effectively zero for a vacuum state. This sets the minimum uncertainty (the shot-noise level) to a value of 1, which is a consequence of the chosen quadrature definitions in Eq.(\ref{1}).

A similar picture may be attributed to that of a single two-level emitter such that:
\begin{align}
\label{6}
\langle\Delta U_{\phi}^{2}\rangle&=2\Big(\langle\Delta \sigma^\dagger\Delta\sigma\rangle +\Re\Big\{e^{-i2\phi}\langle\Delta \sigma^2\rangle\Big\}\Big){-}\langle\sigma_z\rangle.
\end{align} 
This resulting expression can be understood in analogy with Eq.(\ref{5}), consisting of a dipole transition's coherent and incoherent contributions. However, instead of a constant uncertainty level like that of the optical field, the fluctuation are limited by the instantaneous population inversion $\langle \sigma_z\rangle$, which is a function of the external driving fields and intrinsic decay rates. 

To obtain a consistent prognosis, $\langle\sigma_z\rangle$ is set to -$1/2$, rather then being discarded via employing normal-ordering during the derivation of Eq.(\ref{6}). This corresponds to the minimum value it takes when the coherent fluctuations $\langle \Delta\sigma^2\rangle$ are at a maximum, and for which the minimum uncertainty product abides by $\big\langle\Delta U_{\phi}^{2}\big\rangle\big\langle\Delta U_{\phi + \frac{\pi}{2}}^{2}\big\rangle{\geq}1/4$. 

Considering an ensemble of non-interacting dipoles, the field-mediated intra-ensemble fluctuations is linearly summed, rather then summed in quadrature, as all the individual fluctuations are correlated via their identical coupling to the same cavity: 
\begin{align}
\label{7}
\Sigma_{inc} &= N\langle\Delta \sigma_1^\dagger\Delta\sigma_1\rangle + (N{-}1) \langle\Delta \sigma_1^\dagger\Delta\sigma_2\rangle, \nonumber
\\
\Sigma_{coh} &= N\langle\Delta \sigma_1^2\rangle + (N{-}1)\langle\Delta \sigma_1\Delta\sigma_2\rangle, \nonumber
\\
\langle\Delta U_{\phi}^{2}\rangle'&=2\Big(\Sigma_{inc} +\Re\big\{e^{-i2\phi}\Sigma_{coh}\big\}\Big){+}\frac{N}{2}.
\end{align}

The degree of quadrature squeezing is conventionally characterized using homodyne detection, where the signal of interest is mixed with a local oscillator at a given phase $\theta$ and a frequency $\omega_{_{LO}}$, generating sidebands at the frequencies $\omega_{_{LO}} \pm \nu$. Upon detection with a suitable bandwidth detector, these are converted into a low-frequency photocurrent, whose spectral density $S_\theta(\nu)$ directly measures the $\nu$-dependent noise variance. 

Given the weak-sense stationary nature of the rate equations, $S_\theta(\nu)$ is by definition the Fourier transform of the field quadratures auto-correlation function (\textit{g}$^{(1)}$), via the Wiener-Khintchine theorem. However, in the case where an analytical expression is not sought, it is numerically convenient to directly estimate the spectral density of the integrated rate equations using a periodogram-based computation (e.g. Welch's method \cite{Welch1967}).  

The periodogram, $\hat{P}$, of the normally ordered variance, $\langle{:}\Delta X_{\theta}^{2}(t){:}\rangle $ (which excludes the constant term), is scaled as a product of the collection and detection efficiencies $\eta_d$ (the probability of collecting one photon and generating one photoelectron) and the cavity escape efficiency $\eta_\kappa$ [$ = \kappa/(\kappa + nonradiative~loss~rate)$ ] to obtain an estimate of $S_\theta(\nu)$: 
\begin{align}
\label{8}
S_\theta(\nu) &=  1 + \eta_d\eta_\kappa \hat{P}\Big\{\langle{:}\Delta X_{\theta}^{2}(t){:}\rangle\Big\}^{1/2}
\end{align}
For constant steady-state quadrature variances, a power spectrum is not meaningful as there are no frequency components other than what is introduced. The power spectrum is therefore only calculated for oscillating solutions, which directly conveys the distribution and magnitude of the quadrature fluctuations frequency components. 
 
\subsection{Spin s queezing}

The fluctuations of an ensemble of two-level states are conventionally assessed via a collective angular momentum operator $\langle \bm{J} \rangle = \{\langle J_x\rangle,\langle J_y\rangle,\langle J_z\rangle\}$, also referred to as the collective spin. While there are a few definitions of spin-squeezing depending on the experimental settings and the observables of interest \cite{Hammerer2010, Ma2011}, the variance and basic uncertainty relationships of the collective spin components is defined in terms of the three orthogonal coordinates of the collective state Bloch sphere:    
\begin{align}
\label{9}
\langle J_\ell\rangle = \sum^N_k\frac{\langle\sigma_{k\ell}\rangle}{2}
\end{align}
where $\ell = \{x,y,z\}$ designates the given Pauli matrix, which are defined in term of the pseudo-spin operators. Their variance, and the basic uncertainty relation, are further defined with respect to the previously described symmetric treatment of the ensemble: 
\begin{align}
\label{10}
\langle \Delta J_\ell^2 \rangle &= \frac{N}{4}\Big(\langle \sigma_{1\ell} ^2\rangle + (N{-}1)\langle \sigma_{1\ell} \sigma_{2\ell}\rangle\Big) - \langle J_\ell\rangle^2, 
\\
\label{11}
\langle \Delta J_j^2 &\rangle < \tfrac{1}{4}\sqrt{\langle J_k\rangle^2 + \langle J_l\rangle^2},
\end{align}
where $\{j,k,l\}$ represent the three orthogonal spin-coordinates. 

For this uncertainty relationship, it is possible for squeezing to occur simultaneously in the two orthogonal directions of the Bloch sphere's equatorial plane  (see Fig.\ref{fig1}), generating \textit{planar} spin-squeezed states as opposed to standard squeezed state along only one of the orientations ($J_{x,y}$), and Dicke spin-squeezed states which represent un-polarized ensembles where only the spin coordinate in the axial plane ($J_z$) is squeezed \cite{Pezze2018,Vitagliano2018}. Planar spin-squeezed states are particularly interesting as they enable the simultaneous measurement of imparted phase and amplitude changes beyond the classical limit, unlike their standard counterpart \cite{He2011}. 

It has been established that spin-squeezing directly implies multipartite entanglement, which is deemed metrologicaly useful \cite{Pezze2018} when below a size-dependent threshold \cite{Sorensen2001,Toth2009}: 
\begin{align}
\label{12}
\xi^2_j \equiv \frac{\langle \Delta J_j^2\rangle}{\langle J_k \rangle^2 + \langle J_l \rangle^2} < \frac{1}{N}, 
\end{align} 
Comparing Eq.(\ref{12}) and Eq.(\ref{11}), it is evident that multipartite-entanglement and spin-squeezing are not necessarily correspondent \cite{Ma2011}, however any degree of spin-squeezing immediately implies some magnitude of multipartite entanglement \cite{Sorensen2001a}.

\subsection{Solid-state ensemble densities \& coupling strengths}
 
Considering solid-sate ensembles, the symmetric description of a two-level ensemble employed here implies a uniformity which counters the typical in-homogeneity associated with such systems. As the particular spectral information is not sought here, the specifics of the inhomogeneous distribution is not needed for the following analysis; the individual coupling strengths are considered identical, while the direct effect of density-dependent spectral and pure-dephasing inhomogineity can be crudely accounted for to first order by setting $\gamma_2^* \gg \gamma_1$.

Generally, the cavity coupling strength is proportional to an effective mode volume $V_{ef}$ and scaled by the relative alignment of the transition-moment and the resonant field $\zeta$, such that a collective coupling strength  $\mathcal{G}$ can be defined  as proportional to a given ensemble density $N_d$ and longitudinal decay rate $\gamma_1$.  For optical dipoles, this may be expressed via:  
\begin{align}
\label{13}
\mathcal{G}=\sqrt{N}g_1 \propto \zeta\bigg[\big(N_d V_{ef}\big)\bigg(\frac{3\pi c^3\gamma_1}{2\omega_k^2n^3 V_{ef}}\bigg)\bigg]^{1/2}, 
\end{align}
where $c$ is the speed of light and $n$ is the refractive index of the cavity medium. Based on this, the relationship between $g_1$ and $V_{ef}$ may be considered constant for any given $N_d$. Instead, the allowed values of $g_1$ and $N$ may be delineated for any given $N_d$ with respect to the considered cavity system, and the type of ensemble used. 

\begin{figure}
	\centering
	\includegraphics[scale = 0.65]{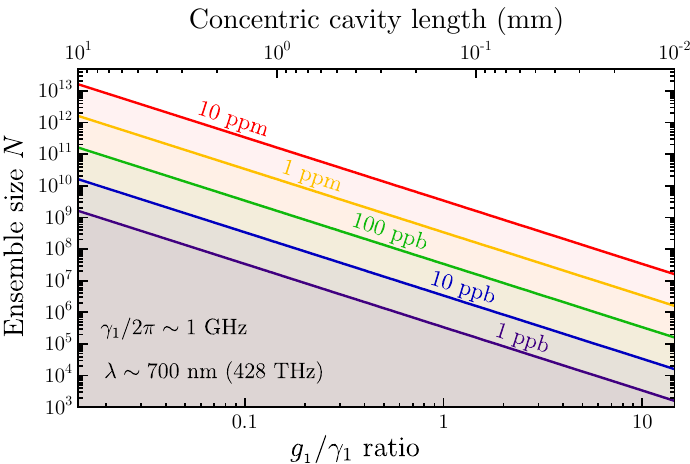}
	\caption{Conceptual limit of the ensemble size for single coupling strengths as a function of varying effective cavity volume. A simple confocal cavity is considered here, where the effective mode volume $V_{ef} = \pi L W_0^2/4$ is varied by increasing its length and the related mirror curvatures, in terms of a near-concentric cavity beam waist defined via $W_0 \approx \sqrt{L\lambda/2\pi}$}
	\label{fig2}
\end{figure}

Considering the simple case of a near-concentric optical cavity, where the mode volume is estimated as a product of a zeroth-order Laguerre-Gaussian beam-waist and the cavity length, the resulting ensemble sizes and expected single emitter coupling strength are plotted in Fig.\ref{fig2} for a range of concentrations. A choice of appropriate coupling rates and ensemble sizes can thereby be based on experimentally determined $\gamma_1$.

For single two-level systems the experimentally achieved coupling strength has typically been four to six orders of magnitude lower than the transition frequency \cite{FriskKockum2019}, while the ratio $g_1/\{\gamma_1,\kappa\}$ can span between 0.1 to 100 for highly optimized systems, but are usually two to three orders of magnitude lower than $\gamma_1$.     

For solid-state optical defects such as tin vacancy centers in diamond, the average decay rates in the order of $200$ MHz have been measured from ensembles with densities estimated in the order of 1 ppm \cite{Haussler2017,Iwasaki2017}. Alternatively, for rare-earth ion systems such as europium-doped yttrium silicate or praseodymium-doped yttrium aluminum garnet, decay rates down to $100$ Hz below $10$ K, and up to $50$ MHz at room temperature have been measured \cite{Yano1992,Kolesov2012}.
 
Further accounting for the crystal symmetry e.g. the tetrahedral symmetry of diamond and how the tin vacancy ensemble's dipole orientations will be distributed over four distinct orientations, the dipole alignment factor $\zeta$ ranges between 0.5 and 0.75, such that a realistic collective coupling strength can be considered to range from $\mathcal{G} \propto 10^{-8}\omega_k$ for diamond-based ensembles, and up to $\propto 10^{-3}\omega_k$ for rare-earth ion ensembles, considering the rough scaling in Fig.\ref{fig2}.  

Conceptually, the proportion between $N$ and $g_1$ can be modified by varying the concentration, but the issues associated with larger concentrations are nontrivial. As well as increased inhomogeneous broadening, a hard limit on the feasible density exists, beyond which the defect loses its integrity and desired transition properties. 

However, as demonstrated using dense rare-earth ion ensembles, inhomogeneous broadening can be circumvented via spectral-hole burning techniques, e.g. \cite{Nilsson2004}. Ideally, a system where the ground- and excited-state hyperfine transition frequencies exceed the inhomogeneous broadening frequency is desirable, such as Ho$^{3+}$ \cite{Agladze1991}, in order to avoid issues related to the modification of the cavity's free spectral range and decay rate by the hole-burning procedure \cite{Shakhmuratov2005,Sabooni2013}. 

Another possibility could involve preparing highly concentrated colloidal quantum dot aggregates. Such systems characteristically possess much faster decoherence and longitudinal decay rates, but provide the advantage of facilitating the creation of comparably homogeneous concentrations exceeding 100 ppm, which span larger volumes, with the appealing potential for wavelength tune-ability by adjusting their size. Considering a recent example demonstrating discrete single photon emission from colloidal perovskite-based quantum dots \cite{Utzat2019}, the longitudinal decay rates are measured to be an order of magnitude faster then those for diamond defects, which projects possible rates in the order of $\mathcal{G}\propto10^{-4}\omega_k$, considering the example scaling in Fig.\ref{fig2}. 

\section{Continuously Driven Squeezing} \label{Sec3}
\subsection{Single dipole}

For the case of a single emitter without a cavity ($N=1,~g_1=0$), a direct analytical solution may be obtained for the steady-state quadrature fluctuations, in a rotating frame with the driving field frequency $\omega_l$. The steady-state expression in terms of a scaled Rabi frequency $z=(\Omega/\vert\Gamma_t\vert)^2$ is derived as: 
\begin{align}
\label{14}
&\langle \Delta U_{\phi}^{2} \rangle_s' = \frac{z\alpha}{(2z\alpha + 1)} - \Re\Bigg\{\frac{z(1 + e^{-i2\phi})}{4(2z\alpha + 1)^2}\Bigg\} + \frac{1}{2},
\end{align}
where $\Gamma_t{=}\Gamma{+}i\Delta_0$, $\alpha{=}\Gamma/2\gamma_1$, $\Gamma{=}\gamma_1/2{+}\gamma_2^*{+}\bar{n}\gamma_1$ is the total temperature-dependent dephasing rate, $\Delta_0{=}(\omega_0{-}\omega_l)$ is the detuning with respect to the driving field, and the subscript $s$ denotes a settled steady-state after a duration such that $ \tfrac{d}{dt} \langle...\rangle = 0$.

\begin{figure}
	\centering
	\includegraphics[scale = 0.87]{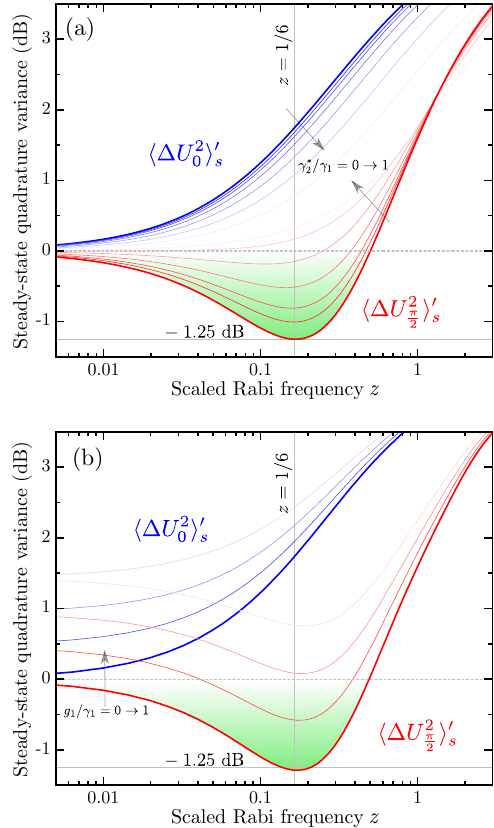}
	\caption{(a) Quadrature fluctuations of a free-space single emitter as a function of the scaled Rabi frequency $z$ for various dephasing rates $\gamma_2^*/\gamma_1 = \{0,0.1,..1 \}$, and (b) for various cavity-coupling rates $g_1/\gamma_1=\{0,0.4,0.6,1\}$, where $\kappa/\gamma_1 = 10$. Blue and red traces represent orthogonal quadrature variances.}
	\label{fig3}
\end{figure}

Neglecting heat ($\bar{n} = 0$), Eq.(\ref{14}) is plotted in Fig.\ref{fig3}(a), and shows that the minimum squeezed variance is obtained for $z  = 1/6$, in the order of ${-}1.25$ dB. Introducing dephasing drastically reduces the difference between the two orthogonal quadratures, such that no squeezing may be generated when $\gamma_2^*>\gamma_1$. 

When coupling a single dipole to a cavity and coherently driving its transition resonantly ($\Delta_c{=}(\omega_c{-}\omega_l){=}0{=}\Delta_0$) using an off-axis field, a simultaneous increase in the fluctuations of both orthogonal quadratures is generated as $\Omega$ is increased, which is plotted in Fig.\ref{fig3}(b). 

\begin{figure}
	\includegraphics[scale = 3.5]{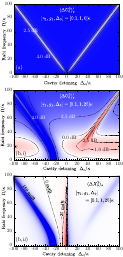}
	\centering
	\caption{Steady-state map of the minimum quadrature variance as a function of the cavity detuning $\Delta_c$ and the Rabi frequency $\Omega$, for (a) resonant and (b) detuned single emitter and driving field.}
	\label{fig4}
\end{figure}

Based on a cursory analysis of the steady-state form of the coupled rate equations (see appendix), this may be considered a consequence of the enhanced exchange rate of quanta between the cavity field and the dipole $\langle a^\dagger\sigma_1\rangle_s$, which leads to the simultaneous reduction of the coherent fluctuations $\langle \Delta\sigma_1^2 \rangle_s$, and the increase of incoherent fluctuations $\langle \Delta\sigma_1^\dagger\Delta\sigma_1 \rangle_s$: 
\begin{align}
\label{15}
\langle \Delta\sigma_1^2 \rangle_s  &\propto -\frac{g_1^2}{\Gamma_t^2}\langle a \sigma_{1z}\rangle_s^2,\nonumber
\\
\langle \Delta\sigma_1^\dagger\Delta\sigma_1 \rangle_s &\propto \frac{g_1}{\gamma_1}\langle a^\dagger\sigma_1\rangle_s.
\end{align}

Evidently, these counteractive mechanisms may be mitigated by detuning the dipole from the driving frequency, in particular towards $\Delta_0 > g_1$ which reduces the coherence-reducing correlation $\langle a\sigma_{1z}\rangle_s$ quadratically compared to the detuning-insensitive (to first-order) exchange of quanta $\langle a^\dagger \sigma_1 \rangle_s$.

\begin{figure}	
	\includegraphics[scale = 0.9]{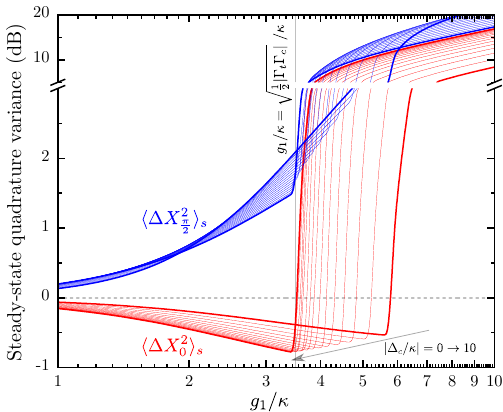}
	\centering
	\caption{Plot of the cavity field steady-state quadrature variance as a function of coupling strengths $g_1$ for a single emitter for various cavity field detunings. The parameters used are $\Omega/\kappa = 1$, $\gamma_1/\kappa = 0.1$, and $\Delta_0/\kappa = 25$, as those used for the detuned case in Fig.\ref{4}(b).}
	\label{fig5}
\end{figure}

The resulting enhancement is demonstrated in Fig.\ref{4} which compares both resonant and detuned configurations. In the resonant case, the quadrature fluctuations can be seen to be reduced when the cavity is on resonance with the dressed-states generated by the coherent driving field. In the detuned case, squeezing in the order of -2.5 dB may be achieved when the dressed state (generalized Rabi) frequency matches that of the detuning frequency, such that $\Delta_c \approx \sqrt{\Omega^2 + \Delta_0^2}$. These relationships have been established through the work of e.g. \cite{Quang1994,Grunwald2013}, and presents an exploitable link between the ensemble's far-field quadrature and the cavity output's quadrature, which is plotted in Fig.\ref{fig4}(b,\textit{ii}). 

Compellingly, the quadrature variance of the cavity output is also modified, showing non-negligible squeezing in the anti-detuned case. Further analysis of the steady-state expressions points towards a mechanism based on the detuning-dependent relationship between the intra-cavity coherence and the dipole coherence: 
\begin{align}
\label{16}
\langle \Delta a^2 \rangle_s \propto &\frac{g_1N}{\Gamma_c}\langle a \sigma_{1z}\rangle_s -\frac{g_1^2N^2}{\Gamma_c^2} \langle \Delta \sigma^2_1 \rangle_s\nonumber
\\
\Longrightarrow \propto &\frac{g_1N}{\Gamma_c}\langle a \sigma_{1z}\rangle_s +\frac{g_1^4N^2}{\Gamma_c^2\Gamma_t^2} \langle a \sigma_{1z}\rangle_s^2.
\end{align}
This highlights how the correlation between the intra-cavity field and the population inversion $\langle a\sigma_z\rangle_s$ is strongly enhanced or suppressed when the relative detunings are of opposite signs, such that the real and imaginary components of the denominator increase and decrease, respectively. 

Physically, this illustrates how a coherent side-band-driven process of the coupled system generates coupled photons without incoherently populating the cavity (via $\langle a^\dagger a\rangle_s$ and thereby $\langle\Delta a^\dagger\Delta a\rangle_s$). A rough proportionality may thus be defined between the ensemble and cavity quadrature fluctuations such that: 
\begin{align}
\label{17}
\langle \Delta X^2_0 \rangle_s &\propto \bigg(\frac{g_1^2 N}{\Gamma_c\Gamma_t}\bigg)^{2} \langle \Delta U^2_{\pi/2} \rangle'_s,~\text{for}~\kappa < \Delta_c.
\end{align}
This indicates how, for low driving rates such that $ \langle \Delta U^2_{\pi/2} \rangle'_s$ is squeezed [cf. Eq.(\ref{14})] and at detunings beyond the cavity decay rate, matching the product of oppositely signed cavity and emitter detuning to the square of the coupling strength [i.e. the product of the denominator in Eq.(\ref{17}) is maximized for $\Im\{\Gamma_c\}{<}0$, $\Im\{\Gamma_t\}{>}0$] to exceed the numerator), any squeezing generated in the ensemble can be proportionally transferred to the cavity field. 

This relation can be understood in terms of how the direct exchange of quanta via the coherent coupling term can be regulated by compensating for the difference between the cavity and two-level relaxation rates through the relative detuning. This resulting squeezing is thereby measurable in the cavity output field, and enhanced by appropriately set relative detunings to offset larger coupling rates. 

Going further, the generated virtual dressed state via the detuned-driving, in addition to the direct transition coupled to the cavity, can be understood to constitute a three-level scheme. This can thereby facilitate lasing beyond both a given coupling or driving rate, which manifests as an exponential increase in both the cavity occupation $\langle a ^\dagger a \rangle_s$ and quadrature variances. This is demonstrated in Fig.\ref{fig5} as a function of $g_1$ for varying values of $\Delta_c$.

The onset of lasing occurs prominently in a far-detuned regime ($\kappa < 2\Delta_c$) when the coupling exceeds the product of the ensemble and cavity decay rates $g_1^2 N \gtrsim \tfrac{1}{2}\Gamma_c\Gamma_t$. By varying the cavity detuning for a given coupling strength, the system can be tuned to reside just before the threshold where squeezing is optimized. Thus for lower $g_1$ coupling values and larger $\Delta_c$ detuning values, squeezing in the cavity-output quadrature may approach the free-space limit of a single two-level state, at the expense of a reduced cavity-field amplitude.

\subsection{Dipole ensemble}

Increasing the ensemble size leads to an enhancement of all cavity-field related correlations by a factor $N$, which augments the proportionality highlighted in Eq.(\ref{17}). However, this is counteracted by a reducing lasing threshold, beyond which the quadrature variance of both the cavity field and the ensemble far-field increase by an order of magnitude. This driving- and coupling-dependent threshold can however be pushed to higher values at the expense of the cavity field amplitude.    

\begin{figure}	
	\includegraphics[scale = 1.1]{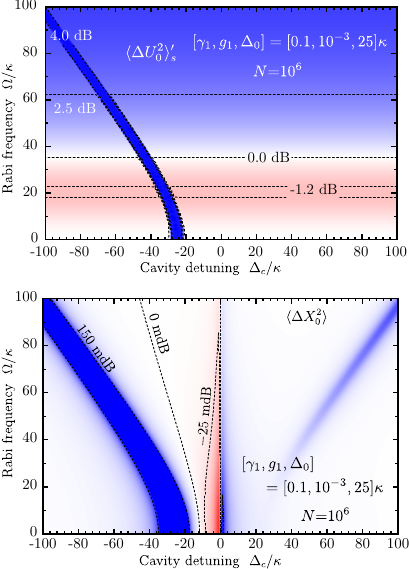}
	\centering
	\caption{Steady-state map of the minimum quadrature variance as a function of cavity detuning $\Delta_c$ and the Rabi frequency $\Omega$, for a detuned ensemble of emitters, using the same parameters as for Fig.\ref{fig4}, except for $g_1$ which is adjusted such that $g_1\sqrt{N}/\kappa = \mathcal{G}/\kappa = 1$.}
	\label{fig6}
\end{figure}

\begin{figure}	
	\includegraphics[scale = 0.8]{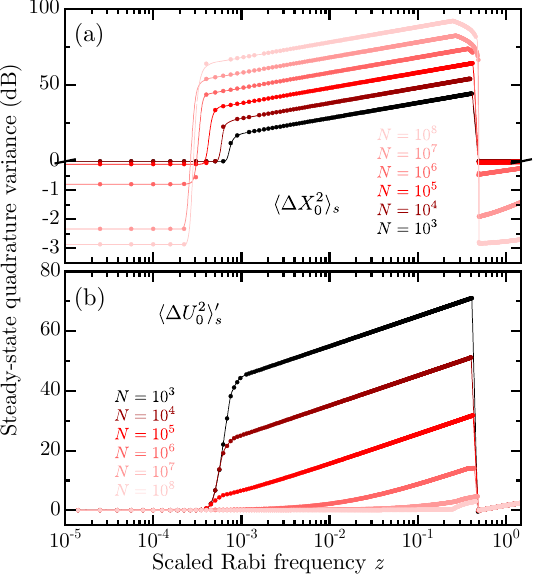}
	\centering
	\caption{Plot of the steady-state quadrature variances as a function of the scaled driving frequency $z$ for for various ensemble sizes. The ensemble detuning here is set to $\Delta_0/\kappa = 80$, while the remaining parameters are kept identical to those used in Fig.\ref{fig6} $\gamma_{1}/\kappa = 0.1$, $\Delta_c/\kappa = -5$, and $\gamma_2^*= 0$.}
	\label{fig7}
\end{figure}

\begin{figure}	
	\includegraphics[scale = 0.85]{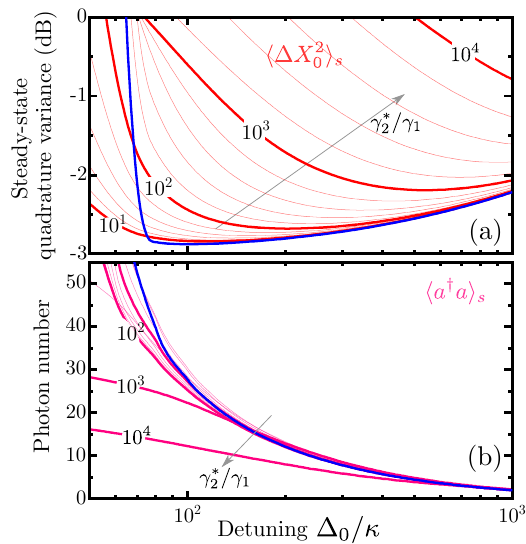}
	\centering
	\caption{Plot of the (a) minimum steady-state quadrature variance of the cavity field and (b) the steady-state photon number of the cavity field, for a range of $\gamma_2^*/\gamma_1$ fractions, which are annotated in the plots. The same parameters as in Fig.\ref{fig7} are used for $N = 10^8$ and $z = 10^{-6}$. The blue traces represent $\gamma_2^* = 0$, while the annotated thick lines are increasing $\gamma_2^*/\gamma_1$ order of magnitude.}
	\label{fig8}
\end{figure}

Replacing the single emitter with an ensemble, the resulting detuning-dependence is plotted in Fig.\ref{fig6}, for which the collective coupling strength is set to equal the value of $g_1$ used in Fig.\ref{fig4}, such that $\mathcal{G}/\kappa = 1$ ($N = 10^6$, $g_1 =10^{-3}$). 

Given the assumption of a non-interacting ensemble, the reduction of $g_1$ implies that $\langle \Delta U_\phi^2\rangle_s$ will resemble that of the single free-space emitter. In this case, provided a weakly-driven regime where $\Omega < 2\Delta_0$, the intra-ensemble correlations are negligible, with the exception of when the cavity is tuned into anti-resonance with the ensemble's driven dressed state ($\Delta_c \approx -\sqrt{\Omega^2 + \Delta_0^2}$). The effect on the cavity output quadrature remains near identical, as expected considering the proportionality described in Eqs.(\ref{16},\ref{17}) . 

The influence of ensemble size is explored in Fig.\ref{fig7}, highlighting how the cavity field may be squeezed towards a limit of -3 dB in the absence of pure-dephasing. In particular, it shows how this is reached by increasing the ensemble size under weak driving. In terms of the proportionality in  Eq.(\ref{17}), this results in a decreased ensemble quadrature variance $\langle \Delta U_\phi^2\rangle_s$ by virtue of the increased coherent intra-ensemble fluctuations $\langle \Delta \sigma_1\Delta \sigma_2\rangle_s$ (Eq.\ref{7}).  

Beyond the threshold, a phase transition occurs pertaining to an increase in the incoherent intra-ensemble fluctuations $\langle \Delta \sigma_1^\dagger \Delta \sigma_2\rangle_s$, up to the point where the scaled Rabi frequency $z$ matches the detuning frequency. Beyond this value, the strength of the driving field exceeds the rate of the enhanced collective process, and the proportionality outlined in Eq.(\ref{17}) is invalidated by higher-order correlations.   

Interestingly, the transition from a conventional lasing character to a more superradiant one is reflected in the relative change of the ensemble quadrature fluctuations - as the ensemble size is increased, the quadrature fluctuations of the cavity increases while the ensemble fluctuations decrease. This is a result of the concurrent increase in both coherent and incoherent intra-ensemble fluctuations, which increases the number of cavity-photons, as the ensemble size and collective coupling rate increases. 

\begin{figure*}	
	\includegraphics[scale = 0.75]{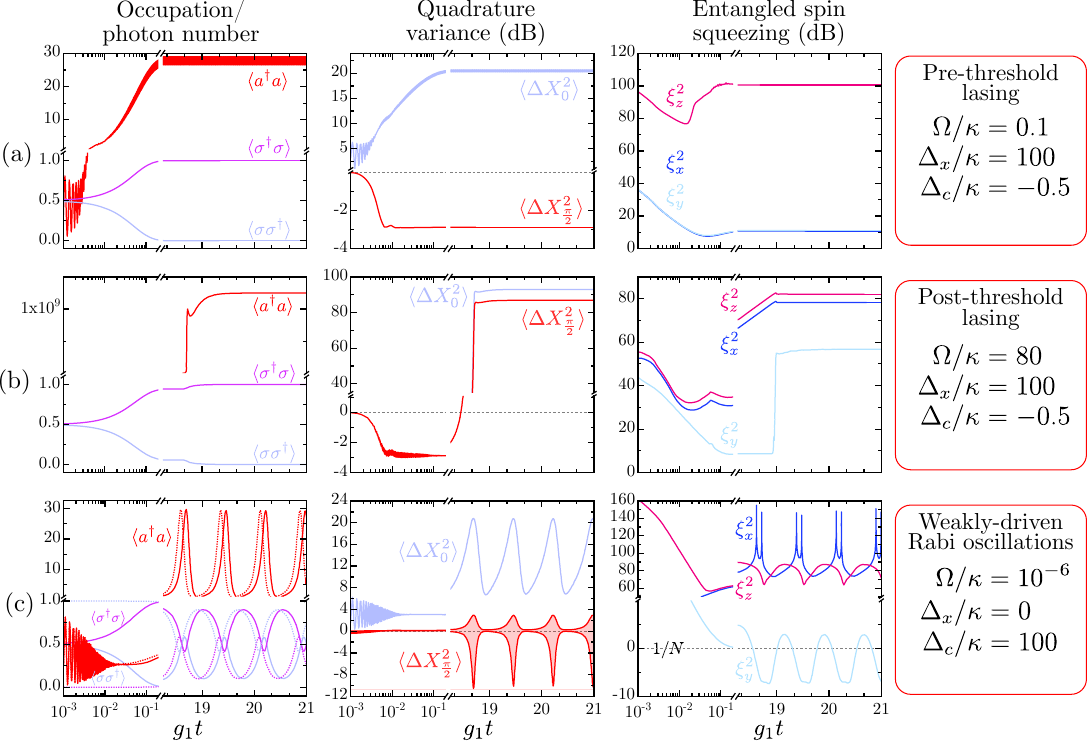}
	\centering
	\caption{Time-dependent dynamics plotted for two different detuning regimes, which demonstrates how quadrature-squeezing may be generated as a steady-state or as a periodic modulation depending on the choice of relative detuning and driving field strength. The dotted lines in the leftmost plot of (c) represent a starting condition of $\langle\sigma_z(0)\rangle = -1$, and is shown to highlight that both thermal and ground-state polarized initial conditions lead to identical periodic behaviour, despite slightly different onsets. Plots simulated for $N = 10^8 $, $g_1/\kappa = 0.005$ and $\gamma_1/\kappa = 0.1$, and $\gamma_2^* = 0$.}
	\label{fig9}
\end{figure*}

Introducing a finite pure-dephasing rate, the degree of squeezing in the cavity-output quadrature is only weakly perturbed, as demonstrated in Fig.\ref{fig8}(a). In particular, the impact of pure-dephasing is observed to be mitigated by varying the cavity detuning, such that by keeping the detuning rate larger then the pure-dephasing rate, irrespective of the cavity and coupling-rate, a degree of squeezing can be maintained.

Provided that the pure-dephasing rate $\gamma_2^*$ does not exceed the ensemble's detuning $\Delta_0$, (such that the proportionality defined in Eq.(\ref{17}) is optimized), the presence of pure-dephasing is therefore not completely detrimental to the generated squeezing of $\langle \Delta X_\theta^2 \rangle_s$. However, this mode of control is useful only so far as the intracavity field retains an experimentally detectable number of photons, shown in Fig.\ref{fig8}(b), which inadvertently decreases $\langle a^\dagger a \rangle_s$ as a function $ \propto 1/\Delta_0$. 

\section{Frequency-modulated quadrature and spin squeezing} \label{Sec4}

Aside from the well-known phenomenon of coherent collapse-and-revival, there are other periodic dynamics, as shown in Fig.\ref{fig9}, which may uniquely generate a degree of quadrature and spin-squeezing. Despite starting from a thermally mixed state, it is possible to generate frequency-modulated quadrature fluctuations, where the periodic enhancement can significantly exceed the optimized steady-state squeezing discussed in the previous section. 

Such periodic modulation transposes itself to the ensembles occupation, which also results in a modulation of the collective angular momenta, via the relationship between the intra-ensemble correlations and the modulated exchange-rate of quanta between the cavity field and the ensemble ( $\langle a^\dagger\sigma_1\rangle$ ). 

Fig.\ref{fig9}(a) shows how a weakly-driven detuned system results in a polarized ensemble which generates a squeezed intra-cavity field. When increasing the driving Rabi frequency $\Omega$ beyond a certain threshold, Fig.\ref{fig9}(b) shows how the system shifts into a lasing superradiant-state, for which the cavity occupation number $\langle a^\dagger a\rangle $ increases exponentially towards the order of $N$, while the ensemble is collectively polarized ($\langle a^\dagger a \rangle/N \geq1$, $ \langle \sigma_1^\dagger\sigma_1 \rangle {-}\langle \sigma_1\sigma_1^\dagger \rangle  = \langle \sigma_{1z} \rangle \simeq 1$). 

When detuning the cavity from the coherent driving frequency (within the cavity bandwidth $\kappa$) towards lower-energies, the intra-cavity field may be indirectly populated at a commensurate rate by a non-resonantly driven ensemble. Conversely, when the cavity is detuned towards higher energies and the ensemble is driven weakly and resonantly, a mixing of the Rabi frequency and the cavity-detuned frequency manifests as a modulation of the cavity quadrature outputs, as shown in Fig.\ref{fig9}(c). 

Despite a thermally mixed starting point, persistent modulation of the intra-cavity field quadratures and the collective angular momenta is generated, for which both the resulting rate and minimum-squeezed magnitude become a function of the driving field frequency and the detuning. In particular, the modulation is comprised of the cavity-detuning frequency enveloped by the much slower coherent driving rate.    

\begin{figure}	
	\includegraphics[scale = 0.65]{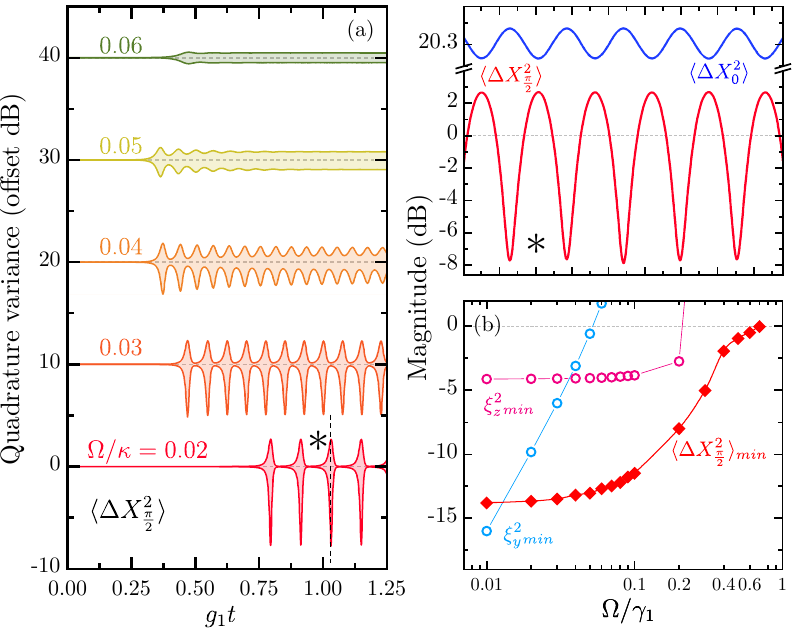}
	\centering
	\caption{(a) Dependence of the dynamic frequency-modulated quadrature on the scaled driving frequency $\Omega/\kappa$, and a magnification of the shaded region (black asterisk) plotting the oscillating quadratures. (b) A plot of the minimum squeezing achieved for the cavity-field quadratures and the maximum degree of spin-squeezed entanglement achieved as a function of driving frequency, for $N = 10^4 $, $g_1/\kappa = 0.005$ and $\gamma_1/\kappa = 0.1$, $\gamma_2^* = 0$, $\Delta_0/\kappa = 0$, and $\Delta_c/\kappa = 100$}
	\label{fig10}
\end{figure}

\begin{figure}	
	\includegraphics[scale = 0.8]{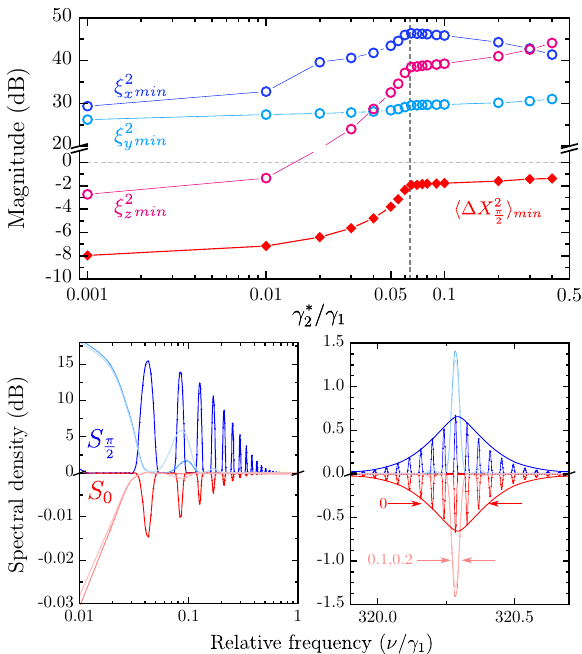}
	\centering
	\caption{Dependence of the minimum frequency-modulated quadrature squeezing and degree of spin-squeezed entanglement on the pure-dephasing rate, plotted using the same parameters as in Fig.\ref{10}, while setting $\Omega/\kappa = 0.02$. The trends highlight two distinct regimes where the dephasing rate exceeds the scaled sum of the coupling and driving frequencies ($\gamma_2^*/\gamma_1 = \big[(g_1/\gamma_1)^2 + (2\Omega/\kappa)^2\big]^{1/2}$, annotated by the vertical dashed line). The spectral densities for two orthogonal quadrature variances are also plotted using Eq.\ref{8} for $\gamma_2^*/\gamma_1 = \{0, 0.1,0.2\}$, and $\eta_d\eta_\kappa = 0.8$.} 
	\label{fig11}
\end{figure}

As shown in Fig.\ref{fig10}, the lower the driving Rabi frequency $\Omega$ and the larger its difference with the cavity detuning $\Delta_c$, the more pronounced the modulation, albeit occurring at slower rates. This reaches an asymptotic limit in the order of -14 dB, while the simultaneous modulated spin-squeezing transitions from a Dicke-like state where only $\xi^2_z$ is squeezed and $\xi^2_{x,y}$ are highly uncertain, to a more planar-like state with squeezing of both $\xi^2_y$ and $\xi^2_z$. 

This difference in scaling between the phase and population-related spin-squeezing can be understood in relation to how the correlations $\langle\sigma_{1z}\sigma_{2z}\rangle$ and $\langle\sigma_{1(x,y)}\sigma_{2(x,y)}\rangle$ scale with $\Omega$ and the ensemble size $N$. For larger ensembles, lower values of $\Omega$ are required to limit the noise contribution of $\langle\sigma_{1z}\sigma_{2z}\rangle$ to $\langle \Delta J_z^2\rangle$ (Eq.[\ref{10}]). Conversely, the phase fluctuations along the orthogonal axis are directly limited by the ratio between the collective coupling rate $\mathcal{G}$ and the detuning of the cavity.

The impact of introducing a finite $\gamma_2^*$, and the resulting spectral density of $\langle \Delta X_\theta ^2\rangle$ is plotted in Fig.\ref{fig11}. Two regimes are delineated around the point where $\gamma_2^*$ equals the sum of the scaled off-axis driving rate and cavity-coupling rate. Unlike the steady-state case, the pure-dephasing rate cannot be mitigated by increasing the cavity detuning. Instead the sum of the individual coupling rate and driving rate need to outcompete $\gamma_2^*$, to ensure that both quadrature and entangled spin-squeezing may be generated.    

The spectral density plots in Fig.\ref{fig11} highlights the well-known property of phase-continuous frequency modulated signals (e.g. sine-wave modulation), which distribute the time dependent amplitude over multiple frequency components, and for which the cumulative integrated power is commensurate with the minimum time-dependent squeezing. Experimentally, data acquisition can be locked at the instances of minimum squeezed variance, which can greatly exceed that of the optimized steady-state value. However, for the small fixed bandwidths within the modulation rate generated here, the spectral density shows how squeezing for single-frequency components will not exceed the free-space limit for these detuning and driving rates. 

\section{Conclusion}

Using a Markov-approximated master equation derived for a Dicke-type Hamiltonian, which describes a cavity-coupled ensemble driven by an off-axis field, the truncated equations of motion (via third-order cumulant expansion) were numerically integrated to explore the generation of quadrature and entangled spin squeezing. The minimum steady-state and frequency-modulated quadrature squeezing was calculated to occur in the limit of -3 dB and -14 dB, respectively, while entangled spin squeezing was calculated to occur at a similar order of magnitude alongside weakly-driven frequency-modulated squeezing.

A direct proportionality between the cavity-field quadrature and the ensembles dipole quadrature was described to scale as a function of the ratio between the collective coupling strength and the relative cavity- and ensemble-loss rates, which is modifiable via the relative detuning of the cavity and the external driving field. Consequentially, the degradation of the cavity-field squeezing by the ensemble's pure-dephasing rate was observed to be mitigated by increasing the relative detuning, at the expense of decreasing the intra-cavity amplitude. 

Frequency-modulated quadrature squeezing was also shown to concur with entangled spin squeezing in a weakly driven regime, where the driving Rabi frequency was orders of magnitude lower than the collective coupling strength. Unlike the steady-state regime, frequency-modulated squeezing is more susceptible to the presence of a finite pure-dephasing rate, which is instead only mitigated by larger cavity coupling rates. 
 
Albeit using a rudimentary Dicke Hamiltonian and Markov-approximated rate equations, this work highlights the possibility of continuously generating quadrature squeezed light from a cavity via applying an off-axis drive to a coupled ensemble, which exceeds the free-space limit of a single two-level emitter. Furthermore, the pure-dephasing rate of the ensemble constituents and, by extension, the ensemble's inhomogeneous broadening, may be mitigated with an appropriately detuned and driven configuration, although optimized squeezing is obtained at the expense of the cavity's output field amplitude and bandwith. Notably, the generation of entangled spin-squeezed states is found to be inherent in a weakly driven regime, but is bandwidth limited depending on the driving and pure-dephasing rates.  

The motivating interest of this work has been in exploring the limits in optimizing quadrature squeezing from an ensemble of emitters, using a cavity and off-axis near-resonant driving. While more sophisticated non-Markovian theoretical approaches need to be considered which account for intra-ensemble interactions, these results provide an informative basis for the experimental exploration of the practical aspects of near-resonance fluorescence based squeezing using solid-state dipole ensembles. This is considered in anticipation of providing solid-state systems emitting at shorter-wavelengths and possessing smaller technical footprints, which can compliment established parametric oscillator sources.   

\begin{acknowledgments}
	This work was partly funded by the Villum Foundation (grant No.17524). Jonas Neergaard Nielsen, Lars Rippe, and Sebastian Horvath are gratefully acknowledged for useful comments and discussions.
\end{acknowledgments}
\appendix*
\section{System Dynamics}
The quadrature fluctuations and steady-state expectation values are obtained by integrating the systems Markov-approximated master equation. Accounting for the cavity decay rate $\kappa$ and the radiative damping in the presence of a heat bath $\bar{n}$ with relaxation and pure-dephasing rates $\gamma_1$ and $\gamma_2^*$, respectively, the master equation and the associated Lindblad operator terms take the form:  
\begin{flalign}
\frac{d}{dt}\rho &= -\frac{i}{\hbar}[(\mathcal{H} + \mathcal{H}_\Omega),\rho] + \mathcal{L} _{\kappa} + \mathcal{L} _{\gamma_{1}}+ \mathcal{L} _{\gamma_{2}^*},&
\\\nonumber
\\
\mathcal{L} _{\kappa}&=\kappa\Big[(\bar{n}{+}1)\big(2 a \rho a^\dagger {-} \{a^\dagger a,\rho\}\big){+}\bar{n}\big(2a^\dagger \rho a - \{a a^\dagger,\rho\}\big)\Big],\nonumber&
\\
\mathcal{L} _{\gamma_{1}}&=\sum_k^N\frac{\gamma_{1,k}}{2}\Big[(\bar{n}{+}1)\big(2\sigma_k \rho \sigma_k^\dagger - \{\sigma_k^\dagger\sigma_k,\rho\}\big)\nonumber&
\\
&+ \bar{n}\big(2\sigma_k^\dagger \rho\sigma_k -\{\sigma_k\sigma_k^\dagger,\rho\}\big)\Big],\nonumber&
\\
\mathcal{L} _{\gamma_{2}^*} &= \sum_k^N\frac{\gamma_{2,k}^*}{2}\big(\sigma_{kz} \rho \sigma_{kz} - \rho\big),  \nonumber&
\\
\bar{n} &=\Big(e^{\hbar\omega_0(k_B  \mathrm{T})^{-1}} - 1\Big)^{-1}.\nonumber&
\end{flalign}
This system is analytically solvable in the case of free-space emitters ($N \geq 1,~g_1=0$), but when $g_1\neq0$, the process of extracting the equations of motion results in an infinite set of successively increasing correlation orders. 

A common strategy for truncation usually involves assuming some form of weakly driven or perturbed systems where $\langle \sigma_z\rangle$  is set to -1 and is assumed to negligibly change. For ensemble systems this is usually accompanied with the Holstein-Primakoff approximation, subsequently enabling the simplification of higher-order correlations by mapping the spin operators onto bosonic operators. These approximations enable the derivation of a closed set of coupled equations, which have been experimentally validated in weakly driven systems, e.g \cite{Ourjoumtsev2011} . 

However, these approximations are not appropriate when accounting for non-negligible amplitudes of near-resonant driving fields. Furthermore for an ensemble, the number of equations quickly increases to an unworkable amount dependent on the ensemble size. 

As carried out in \cite{Meiser2009, Henschel2010}, and described in section \ref{Sec2}.A, the ensemble is described symmetrically to decouple the $N$ dependence of the number of coupled equations of motion. Following this, an alternative strategy is employed to further simplify and reduce the number of coupled equations, based on expanding the correlations in terms of their cumulant expectation values \cite{Kubo1962,Meiser2009}. This avoids any direct restriction of the coupling and driving field strengths, but does assume that the third-order cumulants are negligible, such that: 
\begin{align}
\langle a b c \rangle &\approx  \langle a b \rangle \langle c \rangle + \langle a c \rangle \langle b \rangle + \langle b c \rangle \langle a \rangle -2\langle a\rangle \langle b \rangle \langle c \rangle.
\end{align} 

The truncation of third-order correlations is justified in the regime where the photon fluctuations negligibly affects the ensemble inversion, which is the case when $\{g_1,\Omega\}/\Delta_c {<} \kappa$. This is especially appropriate in the case where the cavity is only populated indirectly via the emission of the ensemble, and its validity was confirmed by comparing the dynamics of $\langle\sigma_z\rangle(t)$ obtained from fourth- and third-order truncated systems. 

Given the time and computational expense of numerically integrating the fourth-order truncated system, third-order truncation was instead used with the justification that it was sufficiently accurate for the parameter space explored in this work.   

The resulting coupled rate equations (not including the conjugate set of equations) are defined below with the complex  loss rates denoted as $\Gamma_{c} =\kappa+i\Delta_c,~	\Gamma_t =\Gamma + i\Delta_0,  ~	\Gamma =\gamma_1(\tfrac{1}{2} + \bar{n}) + \gamma_2^*$: 
\begin{widetext}	
	{\footnotesize \begin{flalign}		
		\tfrac{d}{dt}\langle a\rangle&=-\Gamma_c\langle a \rangle+g_1N\Big(\langle \sigma_1 \rangle - \langle \sigma_1^\dagger \rangle\Big) ,&
		\\
		\tfrac{d}{dt}\langle\sigma_1\rangle &=-\Gamma_t\langle \sigma_1 \rangle+ g_1\Big(\langle a \sigma_{1z}\rangle +\langle a^\dagger \sigma_{1z}\rangle\Big) +\frac{i\Omega}{2}\langle \sigma_{1z}\rangle,&
		\\
		\tfrac{d}{dt}\langle a^2\rangle &= - 2\Gamma_c\langle a^2\rangle + 2g_1N\Big(\langle a\sigma_1\rangle -\langle a \sigma_1^\dagger\rangle\Big),&	
		\\
		\tfrac{d}{dt}\langle a^\dagger a\rangle &= -2\kappa\Big(\langle a^\dagger a \rangle - \bar{n}\Big) + g_1N\Big(\langle a^\dagger\sigma_1 \rangle + \langle a \sigma_1^\dagger \rangle-\langle a \sigma_1\rangle -\langle a^\dagger \sigma_1^\dagger\rangle   \Big),&
		\\
		\tfrac{d}{dt}\langle \sigma_1^\dagger\sigma_1 \rangle &= -\gamma_{1}\Big(\langle \sigma_1^\dagger\sigma_1\rangle+\bar{n}\langle \sigma_{1z}\rangle\Big)-g_1\Big(\langle a^\dagger\sigma_1 \rangle + \langle a \sigma_1^\dagger\rangle +\langle a \sigma_1\rangle +\langle a^\dagger \sigma_1^\dagger\rangle \Big) -\frac{i\Omega}{2}\Big(\langle\sigma_1^\dagger\rangle - \langle\sigma_1\rangle\Big),&
		\\ \nonumber
		\\
		\tfrac{d}{dt}\langle \sigma_1 \sigma_2 \rangle &= -2\Gamma_t\langle\sigma_1\sigma_2\rangle +2g_1\Big(\langle a\sigma_1\sigma_{2z}\rangle +\langle a^\dagger\sigma_1\sigma_{2z}\rangle \Big) + i\Omega\langle \sigma_1\sigma_{2z}\rangle,&
		\\
		\tfrac{d}{dt}\langle \sigma_1^\dagger\sigma_2 \rangle &= -2\Gamma\langle \sigma_1^\dagger\sigma_2 \rangle + g_1\Big(\langle a^\dagger\sigma_1\sigma_{2z}\rangle + \langle a\sigma^\dagger_1\sigma_{2z}\rangle +\langle a\sigma_1\sigma_{2z}\rangle + \langle a^\dagger\sigma_1^\dagger\sigma_{2z}\rangle \Big) -\frac{i\Omega}{2}\Big( \langle \sigma_1\sigma_{2z}\rangle - \langle\sigma_1^\dagger\sigma_{2z}\rangle\Big),&
		\\
		\tfrac{d}{dt}\langle \sigma_{1z}\sigma_{2z} \rangle &= -4\gamma_1\Big( \langle \sigma_1^\dagger\sigma_1\sigma_{2z}\rangle + \bar{n}\langle\sigma_{1z}\sigma_{2z} \rangle\Big){-}4g_1\Big( \langle a^\dagger \sigma_1\sigma_{2z} \rangle + \langle a \sigma_1^\dagger\sigma_{2z} \rangle  +\langle a \sigma_1\sigma_{2z}\rangle+\langle a^\dagger \sigma_1^\dagger\sigma_{2z}\rangle \Big)  -i2\Omega\Big( \langle \sigma_1^\dagger\sigma_{2z} \rangle- \langle \sigma_1\sigma_{2z}\rangle \Big),&
		\\
		\tfrac{d}{dt}\langle \sigma_1\sigma_{2z} \rangle &= {-}2\Gamma\langle\sigma_1\sigma_{2z}\rangle {-}2\gamma_1\Big(\langle\sigma_1\sigma_2 \sigma_2^\dagger \rangle + \bar{n}\langle \sigma_1\sigma_{2z} \rangle\Big) {+} g_1\Big(\langle a\sigma_{1z}\sigma_{2z}\rangle +\langle a^\dagger\sigma_{1z}\sigma_{2z}\rangle {-} 2\big[\langle a^\dagger \sigma_1\sigma_2 \rangle + \langle a \sigma_1\sigma_2^\dagger \rangle] +\langle a \sigma_1\sigma_2\rangle + \langle a^\dagger\sigma_1\sigma_2^\dagger\rangle\big] \Big)\nonumber&
		\\ &~~~-\frac{i\Omega}{2}\Big(2\big[\langle\sigma_1\sigma_2^\dagger\rangle - \langle\sigma_1\sigma_2\rangle\big] - \langle \sigma_{1z}\sigma_{2z} \rangle  \Big),&
		\\ \nonumber
		\\
		\tfrac{d}{dt}\langle a \sigma_1 \rangle &= -\Big(\Gamma_t {+} \Gamma_c\Big)\langle a\sigma_1\rangle {+}g_1\Big(\langle a^2\sigma_{1z}\rangle +\langle a^\dagger a\sigma_{1z}\rangle-\langle \sigma_1\sigma_1^\dagger\rangle\Big) +g_1(N{-}1)\Big(\langle \sigma_1\sigma_2\rangle -\langle \sigma_1\sigma_2^\dagger\rangle\Big) +\frac{i\Omega}{2}\langle a \sigma_{1z}\rangle,&
		\\
		\tfrac{d}{dt}\langle a^\dagger \sigma_1 \rangle &= -\Big(\Gamma_t {+} \Gamma_c{^\dagger}\Big)\langle a^\dagger\sigma_1\rangle {+}g_1\Big(  \langle \sigma_1^\dagger\sigma_1 \rangle + \langle a^\dagger a \sigma_{1z}\rangle {+}\langle a^\dagger a^\dagger\sigma_{1z}\rangle\Big)+g_1(N{-}1)\Big(\langle \sigma_1\sigma_2^\dagger \rangle  -\langle \sigma_1\sigma_2\rangle\Big) + \frac{i\Omega}{2} \langle a^\dagger \sigma_{1z}\rangle,&
		\\
		\tfrac{d}{dt}\langle a \sigma_{1z} \rangle &= -\Gamma_c\langle a\sigma_{1z}\rangle -2\gamma_1\Big(\langle a \sigma_1^\dagger\sigma_1 \rangle + \bar{n}\langle a \sigma_{1z}\rangle\Big) -g_1\Big(2\big[\langle a^2\sigma_1\rangle +\langle a^2\sigma_1^\dagger\rangle+\langle a^\dagger a\sigma_1\rangle  + \langle a^\dagger a \sigma_1^\dagger\rangle \big] + \langle\sigma_1\rangle +\langle \sigma_1^\dagger\rangle\Big)\nonumber&
		\\
		&~~~+ g_1(N{-}1)\Big(\langle \sigma_1\sigma_{2z} \rangle  -\langle \sigma_1^\dagger\sigma_2\rangle\Big) - i\Omega\Big(\langle a \sigma_1^\dagger\rangle - \langle a \sigma_1\rangle\Big).&
		\end{flalign} }
\end{widetext}
The employed approximations convert the resulting autonomous differential system of equations from an infinite linear set to a finite non-linear one, which does not always converge to an asymptotically stable solution given the presence of a continuous coherent drive. 

While the global stability of the resulting non-linear system can not be analytically assessed using conventional stability theory, a rudimentary analysis of the linearized Jacobian indicates that the system's equilibrium solutions are at least stable for the physically allowed values ($\langle \sigma_z \rangle \in [-1,1]$ and $\langle a^\dagger a \rangle \in [0,{+}\infty)$ ), or will at least converge to a stable periodic solution. 

On the other hand, the resulting system can become severely numerically stiff for parameter configurations that are difficult to predict. This is especially so when detuning is introduced and the single cooperativity $\mathcal{C} = g_1^2/\kappa\gamma_1 > 1$. 

In light of this, some of the steady-state solutions are double checked by numerically integrating the rate equations. This is carried out to ensure that the derivatives converge to a limit where they settle within the numerical solver's tolerance and remain so for a duration corresponding to at least an order of magnitude longer than the reciprocal of the slowest defined rate.

\bibliography{Bibli}

\begin{thebibliography}{48}%
\makeatletter
\providecommand \@ifxundefined [1]{%
 \@ifx{#1\undefined}
}%
\providecommand \@ifnum [1]{%
 \ifnum #1\expandafter \@firstoftwo
 \else \expandafter \@secondoftwo
 \fi
}%
\providecommand \@ifx [1]{%
 \ifx #1\expandafter \@firstoftwo
 \else \expandafter \@secondoftwo
 \fi
}%
\providecommand \natexlab [1]{#1}%
\providecommand \enquote  [1]{``#1''}%
\providecommand \bibnamefont  [1]{#1}%
\providecommand \bibfnamefont [1]{#1}%
\providecommand \citenamefont [1]{#1}%
\providecommand \href@noop [0]{\@secondoftwo}%
\providecommand \href [0]{\begingroup \@sanitize@url \@href}%
\providecommand \@href[1]{\@@startlink{#1}\@@href}%
\providecommand \@@href[1]{\endgroup#1\@@endlink}%
\providecommand \@sanitize@url [0]{\catcode `\\12\catcode `\$12\catcode
  `\&12\catcode `\#12\catcode `\^12\catcode `\_12\catcode `\%12\relax}%
\providecommand \@@startlink[1]{}%
\providecommand \@@endlink[0]{}%
\providecommand \url  [0]{\begingroup\@sanitize@url \@url }%
\providecommand \@url [1]{\endgroup\@href {#1}{\urlprefix }}%
\providecommand \urlprefix  [0]{URL }%
\providecommand \Eprint [0]{\href }%
\providecommand \doibase [0]{http://dx.doi.org/}%
\providecommand \selectlanguage [0]{\@gobble}%
\providecommand \bibinfo  [0]{\@secondoftwo}%
\providecommand \bibfield  [0]{\@secondoftwo}%
\providecommand \translation [1]{[#1]}%
\providecommand \BibitemOpen [0]{}%
\providecommand \bibitemStop [0]{}%
\providecommand \bibitemNoStop [0]{.\EOS\space}%
\providecommand \EOS [0]{\spacefactor3000\relax}%
\providecommand \BibitemShut  [1]{\csname bibitem#1\endcsname}%
\let\auto@bib@innerbib\@empty
\bibitem [{\citenamefont {Schnabel}(2017)}]{Schnabel2017}%
  \BibitemOpen
  \bibfield  {author} {\bibinfo {author} {\bibfnamefont {R.}~\bibnamefont
  {Schnabel}},\ }\href {\doibase 10.1016/j.physrep.2017.04.001} {\bibfield
  {journal} {\bibinfo  {journal} {Physics Reports}\ }\textbf {\bibinfo {volume}
  {684}},\ \bibinfo {pages} {1} (\bibinfo {year} {2017})}\BibitemShut {NoStop}%
\bibitem [{\citenamefont {Aasi}\ \emph {et~al.}(2013)\citenamefont {Aasi},
  \citenamefont {Abadie},\ and\ \citenamefont {Abbott~et al.}}]{Aasi2013}%
  \BibitemOpen
  \bibfield  {author} {\bibinfo {author} {\bibfnamefont {J.}~\bibnamefont
  {Aasi}}, \bibinfo {author} {\bibfnamefont {J.}~\bibnamefont {Abadie}}, \ and\
  \bibinfo {author} {\bibfnamefont {B.~P.}\ \bibnamefont {Abbott~et al.}},\
  }\href {\doibase 10.1038/nphoton.2013.177} {\bibfield  {journal} {\bibinfo
  {journal} {Nature Photonics}\ }\textbf {\bibinfo {volume} {7}},\ \bibinfo
  {pages} {613} (\bibinfo {year} {2013})}\BibitemShut {NoStop}%
\bibitem [{\citenamefont {Asavanant}\ \emph {et~al.}(2019)\citenamefont
  {Asavanant}, \citenamefont {Shiozawa}, \citenamefont {Yokoyama},
  \citenamefont {Charoensombutamon}, \citenamefont {Emura}, \citenamefont
  {Alexander}, \citenamefont {Takeda}, \citenamefont {Yoshikawa}, \citenamefont
  {Menicucci}, \citenamefont {Yonezawa},\ and\ \citenamefont
  {Furusawa}}]{Asavanant2019}%
  \BibitemOpen
  \bibfield  {author} {\bibinfo {author} {\bibfnamefont {W.}~\bibnamefont
  {Asavanant}}, \bibinfo {author} {\bibfnamefont {Y.}~\bibnamefont {Shiozawa}},
  \bibinfo {author} {\bibfnamefont {S.}~\bibnamefont {Yokoyama}}, \bibinfo
  {author} {\bibfnamefont {B.}~\bibnamefont {Charoensombutamon}}, \bibinfo
  {author} {\bibfnamefont {H.}~\bibnamefont {Emura}}, \bibinfo {author}
  {\bibfnamefont {R.~N.}\ \bibnamefont {Alexander}}, \bibinfo {author}
  {\bibfnamefont {S.}~\bibnamefont {Takeda}}, \bibinfo {author} {\bibfnamefont
  {J.-i.}\ \bibnamefont {Yoshikawa}}, \bibinfo {author} {\bibfnamefont {N.~C.}\
  \bibnamefont {Menicucci}}, \bibinfo {author} {\bibfnamefont {H.}~\bibnamefont
  {Yonezawa}}, \ and\ \bibinfo {author} {\bibfnamefont {A.}~\bibnamefont
  {Furusawa}},\ }\href {\doibase 10.1126/science.aay2645} {\bibfield  {journal}
  {\bibinfo  {journal} {Science}\ }\textbf {\bibinfo {volume} {366}},\ \bibinfo
  {pages} {373} (\bibinfo {year} {2019})}\BibitemShut {NoStop}%
\bibitem [{\citenamefont {Larsen}\ \emph {et~al.}(2019)\citenamefont {Larsen},
  \citenamefont {Guo}, \citenamefont {Breum}, \citenamefont
  {Neergaard-Nielsen},\ and\ \citenamefont {Andersen}}]{Larsen2019}%
  \BibitemOpen
  \bibfield  {author} {\bibinfo {author} {\bibfnamefont {M.~V.}\ \bibnamefont
  {Larsen}}, \bibinfo {author} {\bibfnamefont {X.}~\bibnamefont {Guo}},
  \bibinfo {author} {\bibfnamefont {C.~R.}\ \bibnamefont {Breum}}, \bibinfo
  {author} {\bibfnamefont {J.~S.}\ \bibnamefont {Neergaard-Nielsen}}, \ and\
  \bibinfo {author} {\bibfnamefont {U.~L.}\ \bibnamefont {Andersen}},\ }\href
  {\doibase 10.1126/science.aay4354} {\bibfield  {journal} {\bibinfo  {journal}
  {Science}\ }\textbf {\bibinfo {volume} {366}},\ \bibinfo {pages} {369}
  (\bibinfo {year} {2019})}\BibitemShut {NoStop}%
\bibitem [{\citenamefont {Walshe}\ \emph {et~al.}(2019)\citenamefont {Walshe},
  \citenamefont {Mensen}, \citenamefont {Baragiola},\ and\ \citenamefont
  {Menicucci}}]{Walshe2019}%
  \BibitemOpen
  \bibfield  {author} {\bibinfo {author} {\bibfnamefont {B.~W.}\ \bibnamefont
  {Walshe}}, \bibinfo {author} {\bibfnamefont {L.~J.}\ \bibnamefont {Mensen}},
  \bibinfo {author} {\bibfnamefont {B.~Q.}\ \bibnamefont {Baragiola}}, \ and\
  \bibinfo {author} {\bibfnamefont {N.~C.}\ \bibnamefont {Menicucci}},\ }\href
  {\doibase 10.1103/PhysRevA.100.010301} {\bibfield  {journal} {\bibinfo
  {journal} {Physical Review A}\ }\textbf {\bibinfo {volume} {100}},\ \bibinfo
  {pages} {010301} (\bibinfo {year} {2019})}\BibitemShut {NoStop}%
\bibitem [{\citenamefont {Singh}\ \emph {et~al.}(2019)\citenamefont {Singh},
  \citenamefont {Ast}, \citenamefont {Mehmet}, \citenamefont {Vahlbruch},\ and\
  \citenamefont {Schnabel}}]{Singh2019}%
  \BibitemOpen
  \bibfield  {author} {\bibinfo {author} {\bibfnamefont {A.~P.}\ \bibnamefont
  {Singh}}, \bibinfo {author} {\bibfnamefont {S.}~\bibnamefont {Ast}}, \bibinfo
  {author} {\bibfnamefont {M.}~\bibnamefont {Mehmet}}, \bibinfo {author}
  {\bibfnamefont {H.}~\bibnamefont {Vahlbruch}}, \ and\ \bibinfo {author}
  {\bibfnamefont {R.}~\bibnamefont {Schnabel}},\ }\href {\doibase
  10.1364/OE.27.022408} {\bibfield  {journal} {\bibinfo  {journal} {Optics
  Express}\ }\textbf {\bibinfo {volume} {27}},\ \bibinfo {pages} {22408}
  (\bibinfo {year} {2019})}\BibitemShut {NoStop}%
\bibitem [{\citenamefont {Arnbak}\ \emph {et~al.}(2019)\citenamefont {Arnbak},
  \citenamefont {Jacobsen}, \citenamefont {Andrade}, \citenamefont {Guo},
  \citenamefont {Neergaard-Nielsen}, \citenamefont {Andersen},\ and\
  \citenamefont {Gehring}}]{Arnbak2019}%
  \BibitemOpen
  \bibfield  {author} {\bibinfo {author} {\bibfnamefont {J.}~\bibnamefont
  {Arnbak}}, \bibinfo {author} {\bibfnamefont {C.~S.}\ \bibnamefont
  {Jacobsen}}, \bibinfo {author} {\bibfnamefont {R.~B.}\ \bibnamefont
  {Andrade}}, \bibinfo {author} {\bibfnamefont {X.}~\bibnamefont {Guo}},
  \bibinfo {author} {\bibfnamefont {J.~S.}\ \bibnamefont {Neergaard-Nielsen}},
  \bibinfo {author} {\bibfnamefont {U.~L.}\ \bibnamefont {Andersen}}, \ and\
  \bibinfo {author} {\bibfnamefont {T.}~\bibnamefont {Gehring}},\ }\href
  {\doibase 10.1364/OE.27.037877} {\bibfield  {journal} {\bibinfo  {journal}
  {Optics Express}\ }\textbf {\bibinfo {volume} {27}},\ \bibinfo {pages}
  {37877} (\bibinfo {year} {2019})}\BibitemShut {NoStop}%
\bibitem [{\citenamefont {Michael}\ \emph {et~al.}(2019)\citenamefont
  {Michael}, \citenamefont {Bello}, \citenamefont {Rosenbluh},\ and\
  \citenamefont {Pe'er}}]{Michael2019}%
  \BibitemOpen
  \bibfield  {author} {\bibinfo {author} {\bibfnamefont {Y.}~\bibnamefont
  {Michael}}, \bibinfo {author} {\bibfnamefont {L.}~\bibnamefont {Bello}},
  \bibinfo {author} {\bibfnamefont {M.}~\bibnamefont {Rosenbluh}}, \ and\
  \bibinfo {author} {\bibfnamefont {A.}~\bibnamefont {Pe'er}},\ }\href
  {\doibase 10.1038/s41534-019-0197-0} {\bibfield  {journal} {\bibinfo
  {journal} {npj Quantum Information}\ }\textbf {\bibinfo {volume} {5}},\
  \bibinfo {pages} {81} (\bibinfo {year} {2019})}\BibitemShut {NoStop}%
\bibitem [{\citenamefont {de~Andrade}\ \emph {et~al.}(2020)\citenamefont
  {de~Andrade}, \citenamefont {Kerdoncuff}, \citenamefont {Berg-S{\o}rensen},
  \citenamefont {Gehring}, \citenamefont {Lassen},\ and\ \citenamefont
  {Andersen}}]{DeAndrade2020}%
  \BibitemOpen
  \bibfield  {author} {\bibinfo {author} {\bibfnamefont {R.~B.}\ \bibnamefont
  {de~Andrade}}, \bibinfo {author} {\bibfnamefont {H.}~\bibnamefont
  {Kerdoncuff}}, \bibinfo {author} {\bibfnamefont {K.}~\bibnamefont
  {Berg-S{\o}rensen}}, \bibinfo {author} {\bibfnamefont {T.}~\bibnamefont
  {Gehring}}, \bibinfo {author} {\bibfnamefont {M.}~\bibnamefont {Lassen}}, \
  and\ \bibinfo {author} {\bibfnamefont {U.~L.}\ \bibnamefont {Andersen}},\
  }\href {\doibase 10.1364/OPTICA.386584} {\bibfield  {journal} {\bibinfo
  {journal} {Optica}\ }\textbf {\bibinfo {volume} {7}},\ \bibinfo {pages} {470}
  (\bibinfo {year} {2020})}\BibitemShut {NoStop}%
\bibitem [{\citenamefont {Andersen}\ \emph {et~al.}(2015)\citenamefont
  {Andersen}, \citenamefont {Neergaard-Nielsen}, \citenamefont {van Loock},\
  and\ \citenamefont {Furusawa}}]{Andersen2015}%
  \BibitemOpen
  \bibfield  {author} {\bibinfo {author} {\bibfnamefont {U.~L.}\ \bibnamefont
  {Andersen}}, \bibinfo {author} {\bibfnamefont {J.~S.}\ \bibnamefont
  {Neergaard-Nielsen}}, \bibinfo {author} {\bibfnamefont {P.}~\bibnamefont {van
  Loock}}, \ and\ \bibinfo {author} {\bibfnamefont {A.}~\bibnamefont
  {Furusawa}},\ }\href {\doibase 10.1038/nphys3410} {\bibfield  {journal}
  {\bibinfo  {journal} {Nature Physics}\ }\textbf {\bibinfo {volume} {11}},\
  \bibinfo {pages} {713} (\bibinfo {year} {2015})}\BibitemShut {NoStop}%
\bibitem [{\citenamefont {Degen}\ \emph {et~al.}(2017)\citenamefont {Degen},
  \citenamefont {Reinhard},\ and\ \citenamefont {Cappellaro}}]{Degen2017}%
  \BibitemOpen
  \bibfield  {author} {\bibinfo {author} {\bibfnamefont {C.~L.}\ \bibnamefont
  {Degen}}, \bibinfo {author} {\bibfnamefont {F.}~\bibnamefont {Reinhard}}, \
  and\ \bibinfo {author} {\bibfnamefont {P.}~\bibnamefont {Cappellaro}},\
  }\href {\doibase 10.1103/RevModPhys.89.035002} {\bibfield  {journal}
  {\bibinfo  {journal} {Reviews of Modern Physics}\ }\textbf {\bibinfo {volume}
  {89}},\ \bibinfo {pages} {035002} (\bibinfo {year} {2017})}\BibitemShut
  {NoStop}%
\bibitem [{\citenamefont {Vahlbruch}\ \emph {et~al.}(2016)\citenamefont
  {Vahlbruch}, \citenamefont {Mehmet}, \citenamefont {Danzmann},\ and\
  \citenamefont {Schnabel}}]{Vahlbruch2016}%
  \BibitemOpen
  \bibfield  {author} {\bibinfo {author} {\bibfnamefont {H.}~\bibnamefont
  {Vahlbruch}}, \bibinfo {author} {\bibfnamefont {M.}~\bibnamefont {Mehmet}},
  \bibinfo {author} {\bibfnamefont {K.}~\bibnamefont {Danzmann}}, \ and\
  \bibinfo {author} {\bibfnamefont {R.}~\bibnamefont {Schnabel}},\ }\href
  {\doibase 10.1103/PhysRevLett.117.110801} {\bibfield  {journal} {\bibinfo
  {journal} {Physical Review Letters}\ }\textbf {\bibinfo {volume} {117}},\
  \bibinfo {pages} {110801} (\bibinfo {year} {2016})}\BibitemShut {NoStop}%
\bibitem [{\citenamefont {Hoff}\ \emph {et~al.}(2015)\citenamefont {Hoff},
  \citenamefont {Nielsen},\ and\ \citenamefont {Andersen}}]{Hoff2015}%
  \BibitemOpen
  \bibfield  {author} {\bibinfo {author} {\bibfnamefont {U.~B.}\ \bibnamefont
  {Hoff}}, \bibinfo {author} {\bibfnamefont {B.~M.}\ \bibnamefont {Nielsen}}, \
  and\ \bibinfo {author} {\bibfnamefont {U.~L.}\ \bibnamefont {Andersen}},\
  }\href {\doibase 10.1364/OE.23.012013} {\bibfield  {journal} {\bibinfo
  {journal} {Optics Express}\ }\textbf {\bibinfo {volume} {23}},\ \bibinfo
  {pages} {12013} (\bibinfo {year} {2015})}\BibitemShut {NoStop}%
\bibitem [{\citenamefont {Vaidya}\ \emph {et~al.}(2020)\citenamefont {Vaidya},
  \citenamefont {Morrison}, \citenamefont {Helt}, \citenamefont {Shahrokshahi},
  \citenamefont {Mahler}, \citenamefont {Collins}, \citenamefont {Tan},
  \citenamefont {Lavoie}, \citenamefont {Repingon}, \citenamefont {Menotti},
  \citenamefont {Quesada}, \citenamefont {Pooser}, \citenamefont {Lita},
  \citenamefont {Gerrits}, \citenamefont {Nam},\ and\ \citenamefont
  {Vernon}}]{Vaidya2020}%
  \BibitemOpen
  \bibfield  {author} {\bibinfo {author} {\bibfnamefont {V.~D.}\ \bibnamefont
  {Vaidya}}, \bibinfo {author} {\bibfnamefont {B.}~\bibnamefont {Morrison}},
  \bibinfo {author} {\bibfnamefont {L.~G.}\ \bibnamefont {Helt}}, \bibinfo
  {author} {\bibfnamefont {R.}~\bibnamefont {Shahrokshahi}}, \bibinfo {author}
  {\bibfnamefont {D.~H.}\ \bibnamefont {Mahler}}, \bibinfo {author}
  {\bibfnamefont {M.~J.}\ \bibnamefont {Collins}}, \bibinfo {author}
  {\bibfnamefont {K.}~\bibnamefont {Tan}}, \bibinfo {author} {\bibfnamefont
  {J.}~\bibnamefont {Lavoie}}, \bibinfo {author} {\bibfnamefont
  {A.}~\bibnamefont {Repingon}}, \bibinfo {author} {\bibfnamefont
  {M.}~\bibnamefont {Menotti}}, \bibinfo {author} {\bibfnamefont
  {N.}~\bibnamefont {Quesada}}, \bibinfo {author} {\bibfnamefont {R.~C.}\
  \bibnamefont {Pooser}}, \bibinfo {author} {\bibfnamefont {A.~E.}\
  \bibnamefont {Lita}}, \bibinfo {author} {\bibfnamefont {T.}~\bibnamefont
  {Gerrits}}, \bibinfo {author} {\bibfnamefont {S.~W.}\ \bibnamefont {Nam}}, \
  and\ \bibinfo {author} {\bibfnamefont {Z.}~\bibnamefont {Vernon}},\ }\href
  {\doibase 10.1126/sciadv.aba9186} {\bibfield  {journal} {\bibinfo  {journal}
  {Science Advances}\ }\textbf {\bibinfo {volume} {6}},\ \bibinfo {pages}
  {eaba9186} (\bibinfo {year} {2020})}\BibitemShut {NoStop}%
\bibitem [{\citenamefont {Walls}\ and\ \citenamefont
  {Zoller}(1981)}]{Walls1981}%
  \BibitemOpen
  \bibfield  {author} {\bibinfo {author} {\bibfnamefont {D.~F.}\ \bibnamefont
  {Walls}}\ and\ \bibinfo {author} {\bibfnamefont {P.}~\bibnamefont {Zoller}},\
  }\href {\doibase 10.1103/PhysRevLett.47.709} {\bibfield  {journal} {\bibinfo
  {journal} {Physical Review Letters}\ }\textbf {\bibinfo {volume} {47}},\
  \bibinfo {pages} {709} (\bibinfo {year} {1981})}\BibitemShut {NoStop}%
\bibitem [{\citenamefont {Ourjoumtsev}\ \emph {et~al.}(2011)\citenamefont
  {Ourjoumtsev}, \citenamefont {Kubanek}, \citenamefont {Koch}, \citenamefont
  {Sames}, \citenamefont {Pinkse}, \citenamefont {Rempe},\ and\ \citenamefont
  {Murr}}]{Ourjoumtsev2011}%
  \BibitemOpen
  \bibfield  {author} {\bibinfo {author} {\bibfnamefont {A.}~\bibnamefont
  {Ourjoumtsev}}, \bibinfo {author} {\bibfnamefont {A.}~\bibnamefont
  {Kubanek}}, \bibinfo {author} {\bibfnamefont {M.}~\bibnamefont {Koch}},
  \bibinfo {author} {\bibfnamefont {C.}~\bibnamefont {Sames}}, \bibinfo
  {author} {\bibfnamefont {P.~W.~H.}\ \bibnamefont {Pinkse}}, \bibinfo {author}
  {\bibfnamefont {G.}~\bibnamefont {Rempe}}, \ and\ \bibinfo {author}
  {\bibfnamefont {K.}~\bibnamefont {Murr}},\ }\href {\doibase
  10.1038/nature10170} {\bibfield  {journal} {\bibinfo  {journal} {Nature}\
  }\textbf {\bibinfo {volume} {474}},\ \bibinfo {pages} {623} (\bibinfo {year}
  {2011})}\BibitemShut {NoStop}%
\bibitem [{\citenamefont {Schulte}\ \emph {et~al.}(2015)\citenamefont
  {Schulte}, \citenamefont {Hansom}, \citenamefont {Jones}, \citenamefont
  {Matthiesen}, \citenamefont {{Le Gall}},\ and\ \citenamefont
  {Atat{\"{u}}re}}]{Schulte2015}%
  \BibitemOpen
  \bibfield  {author} {\bibinfo {author} {\bibfnamefont {C.~H.~H.}\
  \bibnamefont {Schulte}}, \bibinfo {author} {\bibfnamefont {J.}~\bibnamefont
  {Hansom}}, \bibinfo {author} {\bibfnamefont {A.~E.}\ \bibnamefont {Jones}},
  \bibinfo {author} {\bibfnamefont {C.}~\bibnamefont {Matthiesen}}, \bibinfo
  {author} {\bibfnamefont {C.}~\bibnamefont {{Le Gall}}}, \ and\ \bibinfo
  {author} {\bibfnamefont {M.}~\bibnamefont {Atat{\"{u}}re}},\ }\href {\doibase
  10.1038/nature14868} {\bibfield  {journal} {\bibinfo  {journal} {Nature}\
  }\textbf {\bibinfo {volume} {525}},\ \bibinfo {pages} {222} (\bibinfo {year}
  {2015})}\BibitemShut {NoStop}%
\bibitem [{\citenamefont {Vollmer}\ \emph {et~al.}(2014)\citenamefont
  {Vollmer}, \citenamefont {Baune}, \citenamefont {Samblowski}, \citenamefont
  {Eberle}, \citenamefont {H{\"{a}}ndchen}, \citenamefont
  {Fiur{\'{a}}{\v{s}}ek},\ and\ \citenamefont {Schnabel}}]{Vollmer2014}%
  \BibitemOpen
  \bibfield  {author} {\bibinfo {author} {\bibfnamefont {C.~E.}\ \bibnamefont
  {Vollmer}}, \bibinfo {author} {\bibfnamefont {C.}~\bibnamefont {Baune}},
  \bibinfo {author} {\bibfnamefont {A.}~\bibnamefont {Samblowski}}, \bibinfo
  {author} {\bibfnamefont {T.}~\bibnamefont {Eberle}}, \bibinfo {author}
  {\bibfnamefont {V.}~\bibnamefont {H{\"{a}}ndchen}}, \bibinfo {author}
  {\bibfnamefont {J.}~\bibnamefont {Fiur{\'{a}}{\v{s}}ek}}, \ and\ \bibinfo
  {author} {\bibfnamefont {R.}~\bibnamefont {Schnabel}},\ }\href {\doibase
  10.1103/PhysRevLett.112.073602} {\bibfield  {journal} {\bibinfo  {journal}
  {Physical Review Letters}\ }\textbf {\bibinfo {volume} {112}},\ \bibinfo
  {pages} {073602} (\bibinfo {year} {2014})}\BibitemShut {NoStop}%
\bibitem [{\citenamefont {Meystre}\ and\ \citenamefont
  {Zubairy}(1982)}]{Meystre1982}%
  \BibitemOpen
  \bibfield  {author} {\bibinfo {author} {\bibfnamefont {P.}~\bibnamefont
  {Meystre}}\ and\ \bibinfo {author} {\bibfnamefont {M.}~\bibnamefont
  {Zubairy}},\ }\href {\doibase 10.1016/0375-9601(82)90330-9} {\bibfield
  {journal} {\bibinfo  {journal} {Physics Letters A}\ }\textbf {\bibinfo
  {volume} {89}},\ \bibinfo {pages} {390} (\bibinfo {year} {1982})}\BibitemShut
  {NoStop}%
\bibitem [{\citenamefont {Collett}\ \emph {et~al.}(1984)\citenamefont
  {Collett}, \citenamefont {Walls},\ and\ \citenamefont
  {Zoller}}]{Collett1984}%
  \BibitemOpen
  \bibfield  {author} {\bibinfo {author} {\bibfnamefont {M.}~\bibnamefont
  {Collett}}, \bibinfo {author} {\bibfnamefont {D.}~\bibnamefont {Walls}}, \
  and\ \bibinfo {author} {\bibfnamefont {P.}~\bibnamefont {Zoller}},\ }\href
  {\doibase 10.1016/0030-4018(84)90300-6} {\bibfield  {journal} {\bibinfo
  {journal} {Optics Communications}\ }\textbf {\bibinfo {volume} {52}},\
  \bibinfo {pages} {145} (\bibinfo {year} {1984})}\BibitemShut {NoStop}%
\bibitem [{\citenamefont {W{\'{o}}dkiewicz}\ \emph {et~al.}(1987)\citenamefont
  {W{\'{o}}dkiewicz}, \citenamefont {Knight}, \citenamefont {Buckle},\ and\
  \citenamefont {Barnett}}]{Wodkiewicz1987}%
  \BibitemOpen
  \bibfield  {author} {\bibinfo {author} {\bibfnamefont {K.}~\bibnamefont
  {W{\'{o}}dkiewicz}}, \bibinfo {author} {\bibfnamefont {P.~L.}\ \bibnamefont
  {Knight}}, \bibinfo {author} {\bibfnamefont {S.~J.}\ \bibnamefont {Buckle}},
  \ and\ \bibinfo {author} {\bibfnamefont {S.~M.}\ \bibnamefont {Barnett}},\
  }\href {\doibase 10.1103/PhysRevA.35.2567} {\bibfield  {journal} {\bibinfo
  {journal} {Physical Review A}\ }\textbf {\bibinfo {volume} {35}},\ \bibinfo
  {pages} {2567} (\bibinfo {year} {1987})}\BibitemShut {NoStop}%
\bibitem [{\citenamefont {Kitagawa}\ and\ \citenamefont
  {Ueda}(1993)}]{Kitagawa1993}%
  \BibitemOpen
  \bibfield  {author} {\bibinfo {author} {\bibfnamefont {M.}~\bibnamefont
  {Kitagawa}}\ and\ \bibinfo {author} {\bibfnamefont {M.}~\bibnamefont
  {Ueda}},\ }\href {\doibase 10.1103/PhysRevA.47.5138} {\bibfield  {journal}
  {\bibinfo  {journal} {Physical Review A}\ }\textbf {\bibinfo {volume} {47}},\
  \bibinfo {pages} {5138} (\bibinfo {year} {1993})}\BibitemShut {NoStop}%
\bibitem [{\citenamefont {S{\o}rensen}\ \emph {et~al.}(2001)\citenamefont
  {S{\o}rensen}, \citenamefont {Duan}, \citenamefont {Cirac},\ and\
  \citenamefont {Zoller}}]{Sorensen2001}%
  \BibitemOpen
  \bibfield  {author} {\bibinfo {author} {\bibfnamefont {A.}~\bibnamefont
  {S{\o}rensen}}, \bibinfo {author} {\bibfnamefont {L.-M.}\ \bibnamefont
  {Duan}}, \bibinfo {author} {\bibfnamefont {J.~I.}\ \bibnamefont {Cirac}}, \
  and\ \bibinfo {author} {\bibfnamefont {P.}~\bibnamefont {Zoller}},\ }\href
  {\doibase 10.1038/35051038} {\bibfield  {journal} {\bibinfo  {journal}
  {Nature}\ }\textbf {\bibinfo {volume} {409}},\ \bibinfo {pages} {63}
  (\bibinfo {year} {2001})}\BibitemShut {NoStop}%
\bibitem [{\citenamefont {Haakh}\ and\ \citenamefont
  {Mart{\'{i}}n-Cano}(2015)}]{Haakh2015}%
  \BibitemOpen
  \bibfield  {author} {\bibinfo {author} {\bibfnamefont {H.~R.}\ \bibnamefont
  {Haakh}}\ and\ \bibinfo {author} {\bibfnamefont {D.}~\bibnamefont
  {Mart{\'{i}}n-Cano}},\ }\href {\doibase 10.1021/acsphotonics.5b00585}
  {\bibfield  {journal} {\bibinfo  {journal} {ACS Photonics}\ }\textbf
  {\bibinfo {volume} {2}},\ \bibinfo {pages} {1686} (\bibinfo {year}
  {2015})}\BibitemShut {NoStop}%
\bibitem [{\citenamefont {T{\'{o}}th}\ \emph {et~al.}(2009)\citenamefont
  {T{\'{o}}th}, \citenamefont {Knapp}, \citenamefont {G{\"{u}}hne},\ and\
  \citenamefont {Briegel}}]{Toth2009}%
  \BibitemOpen
  \bibfield  {author} {\bibinfo {author} {\bibfnamefont {G.}~\bibnamefont
  {T{\'{o}}th}}, \bibinfo {author} {\bibfnamefont {C.}~\bibnamefont {Knapp}},
  \bibinfo {author} {\bibfnamefont {O.}~\bibnamefont {G{\"{u}}hne}}, \ and\
  \bibinfo {author} {\bibfnamefont {H.~J.}\ \bibnamefont {Briegel}},\ }\href
  {\doibase 10.1103/PhysRevA.79.042334} {\bibfield  {journal} {\bibinfo
  {journal} {Physical Review A}\ }\textbf {\bibinfo {volume} {79}},\ \bibinfo
  {pages} {042334} (\bibinfo {year} {2009})}\BibitemShut {NoStop}%
\bibitem [{\citenamefont {Vitagliano}\ \emph {et~al.}(2018)\citenamefont
  {Vitagliano}, \citenamefont {Colangelo}, \citenamefont {{Martin Ciurana}},
  \citenamefont {Mitchell}, \citenamefont {Sewell},\ and\ \citenamefont
  {T{\'{o}}th}}]{Vitagliano2018}%
  \BibitemOpen
  \bibfield  {author} {\bibinfo {author} {\bibfnamefont {G.}~\bibnamefont
  {Vitagliano}}, \bibinfo {author} {\bibfnamefont {G.}~\bibnamefont
  {Colangelo}}, \bibinfo {author} {\bibfnamefont {F.}~\bibnamefont {{Martin
  Ciurana}}}, \bibinfo {author} {\bibfnamefont {M.~W.}\ \bibnamefont
  {Mitchell}}, \bibinfo {author} {\bibfnamefont {R.~J.}\ \bibnamefont
  {Sewell}}, \ and\ \bibinfo {author} {\bibfnamefont {G.}~\bibnamefont
  {T{\'{o}}th}},\ }\href {\doibase 10.1103/PhysRevA.97.020301} {\bibfield
  {journal} {\bibinfo  {journal} {Physical Review A}\ }\textbf {\bibinfo
  {volume} {97}},\ \bibinfo {pages} {020301} (\bibinfo {year}
  {2018})}\BibitemShut {NoStop}%
\bibitem [{\citenamefont {Gr{\"{u}}nwald}\ and\ \citenamefont
  {Vogel}(2013)}]{Grunwald2013}%
  \BibitemOpen
  \bibfield  {author} {\bibinfo {author} {\bibfnamefont {P.}~\bibnamefont
  {Gr{\"{u}}nwald}}\ and\ \bibinfo {author} {\bibfnamefont {W.}~\bibnamefont
  {Vogel}},\ }\href {\doibase 10.1103/PhysRevA.88.023837} {\bibfield  {journal}
  {\bibinfo  {journal} {Physical Review A}\ }\textbf {\bibinfo {volume} {88}},\
  \bibinfo {pages} {023837} (\bibinfo {year} {2013})}\BibitemShut {NoStop}%
\bibitem [{\citenamefont {Dimer}\ \emph {et~al.}(2007)\citenamefont {Dimer},
  \citenamefont {Estienne}, \citenamefont {Parkins},\ and\ \citenamefont
  {Carmichael}}]{Dimer2007}%
  \BibitemOpen
  \bibfield  {author} {\bibinfo {author} {\bibfnamefont {F.}~\bibnamefont
  {Dimer}}, \bibinfo {author} {\bibfnamefont {B.}~\bibnamefont {Estienne}},
  \bibinfo {author} {\bibfnamefont {A.~S.}\ \bibnamefont {Parkins}}, \ and\
  \bibinfo {author} {\bibfnamefont {H.~J.}\ \bibnamefont {Carmichael}},\ }\href
  {\doibase 10.1103/PhysRevA.75.013804} {\bibfield  {journal} {\bibinfo
  {journal} {Physical Review A}\ }\textbf {\bibinfo {volume} {75}},\ \bibinfo
  {pages} {013804} (\bibinfo {year} {2007})}\BibitemShut {NoStop}%
\bibitem [{\citenamefont {Meiser}\ \emph {et~al.}(2009)\citenamefont {Meiser},
  \citenamefont {Ye}, \citenamefont {Carlson},\ and\ \citenamefont
  {Holland}}]{Meiser2009}%
  \BibitemOpen
  \bibfield  {author} {\bibinfo {author} {\bibfnamefont {D.}~\bibnamefont
  {Meiser}}, \bibinfo {author} {\bibfnamefont {J.}~\bibnamefont {Ye}}, \bibinfo
  {author} {\bibfnamefont {D.~R.}\ \bibnamefont {Carlson}}, \ and\ \bibinfo
  {author} {\bibfnamefont {M.~J.}\ \bibnamefont {Holland}},\ }\href {\doibase
  10.1103/PhysRevLett.102.163601} {\bibfield  {journal} {\bibinfo  {journal}
  {Physical Review Letters}\ }\textbf {\bibinfo {volume} {102}},\ \bibinfo
  {pages} {163601} (\bibinfo {year} {2009})}\BibitemShut {NoStop}%
\bibitem [{\citenamefont {Henschel}\ \emph {et~al.}(2010)\citenamefont
  {Henschel}, \citenamefont {Majer}, \citenamefont {Schmiedmayer},\ and\
  \citenamefont {Ritsch}}]{Henschel2010}%
  \BibitemOpen
  \bibfield  {author} {\bibinfo {author} {\bibfnamefont {K.}~\bibnamefont
  {Henschel}}, \bibinfo {author} {\bibfnamefont {J.}~\bibnamefont {Majer}},
  \bibinfo {author} {\bibfnamefont {J.}~\bibnamefont {Schmiedmayer}}, \ and\
  \bibinfo {author} {\bibfnamefont {H.}~\bibnamefont {Ritsch}},\ }\href
  {\doibase 10.1103/PhysRevA.82.033810} {\bibfield  {journal} {\bibinfo
  {journal} {Physical Review A}\ }\textbf {\bibinfo {volume} {82}},\ \bibinfo
  {pages} {033810} (\bibinfo {year} {2010})}\BibitemShut {NoStop}%
\bibitem [{\citenamefont {Kubo}(1962)}]{Kubo1962}%
  \BibitemOpen
  \bibfield  {author} {\bibinfo {author} {\bibfnamefont {R.}~\bibnamefont
  {Kubo}},\ }\href {\doibase 10.1143/JPSJ.17.1100} {\bibfield  {journal}
  {\bibinfo  {journal} {Journal of the Physical Society of Japan}\ }\textbf
  {\bibinfo {volume} {17}},\ \bibinfo {pages} {1100} (\bibinfo {year}
  {1962})}\BibitemShut {NoStop}%
\bibitem [{\citenamefont {Welch}(1967)}]{Welch1967}%
  \BibitemOpen
  \bibfield  {author} {\bibinfo {author} {\bibfnamefont {P.}~\bibnamefont
  {Welch}},\ }\href {\doibase 10.1109/TAU.1967.1161901} {\bibfield  {journal}
  {\bibinfo  {journal} {IEEE Transactions on Audio and Electroacoustics}\
  }\textbf {\bibinfo {volume} {15}},\ \bibinfo {pages} {70} (\bibinfo {year}
  {1967})}\BibitemShut {NoStop}%
\bibitem [{\citenamefont {Hammerer}\ \emph {et~al.}(2010)\citenamefont
  {Hammerer}, \citenamefont {S{\o}rensen},\ and\ \citenamefont
  {Polzik}}]{Hammerer2010}%
  \BibitemOpen
  \bibfield  {author} {\bibinfo {author} {\bibfnamefont {K.}~\bibnamefont
  {Hammerer}}, \bibinfo {author} {\bibfnamefont {A.~S.}\ \bibnamefont
  {S{\o}rensen}}, \ and\ \bibinfo {author} {\bibfnamefont {E.~S.}\ \bibnamefont
  {Polzik}},\ }\href {\doibase 10.1103/RevModPhys.82.1041} {\bibfield
  {journal} {\bibinfo  {journal} {Reviews of Modern Physics}\ }\textbf
  {\bibinfo {volume} {82}},\ \bibinfo {pages} {1041} (\bibinfo {year}
  {2010})}\BibitemShut {NoStop}%
\bibitem [{\citenamefont {Ma}\ \emph {et~al.}(2011)\citenamefont {Ma},
  \citenamefont {Wang}, \citenamefont {Sun},\ and\ \citenamefont
  {Nori}}]{Ma2011}%
  \BibitemOpen
  \bibfield  {author} {\bibinfo {author} {\bibfnamefont {J.}~\bibnamefont
  {Ma}}, \bibinfo {author} {\bibfnamefont {X.}~\bibnamefont {Wang}}, \bibinfo
  {author} {\bibfnamefont {C.}~\bibnamefont {Sun}}, \ and\ \bibinfo {author}
  {\bibfnamefont {F.}~\bibnamefont {Nori}},\ }\href {\doibase
  10.1016/j.physrep.2011.08.003} {\bibfield  {journal} {\bibinfo  {journal}
  {Physics Reports}\ }\textbf {\bibinfo {volume} {509}},\ \bibinfo {pages} {89}
  (\bibinfo {year} {2011})}\BibitemShut {NoStop}%
\bibitem [{\citenamefont {Pezz{\`{e}}}\ \emph {et~al.}(2018)\citenamefont
  {Pezz{\`{e}}}, \citenamefont {Smerzi}, \citenamefont {Oberthaler},
  \citenamefont {Schmied},\ and\ \citenamefont {Treutlein}}]{Pezze2018}%
  \BibitemOpen
  \bibfield  {author} {\bibinfo {author} {\bibfnamefont {L.}~\bibnamefont
  {Pezz{\`{e}}}}, \bibinfo {author} {\bibfnamefont {A.}~\bibnamefont {Smerzi}},
  \bibinfo {author} {\bibfnamefont {M.~K.}\ \bibnamefont {Oberthaler}},
  \bibinfo {author} {\bibfnamefont {R.}~\bibnamefont {Schmied}}, \ and\
  \bibinfo {author} {\bibfnamefont {P.}~\bibnamefont {Treutlein}},\ }\href
  {\doibase 10.1103/RevModPhys.90.035005} {\bibfield  {journal} {\bibinfo
  {journal} {Reviews of Modern Physics}\ }\textbf {\bibinfo {volume} {90}},\
  \bibinfo {pages} {035005} (\bibinfo {year} {2018})}\BibitemShut {NoStop}%
\bibitem [{\citenamefont {He}\ \emph {et~al.}(2011)\citenamefont {He},
  \citenamefont {Peng}, \citenamefont {Drummond},\ and\ \citenamefont
  {Reid}}]{He2011}%
  \BibitemOpen
  \bibfield  {author} {\bibinfo {author} {\bibfnamefont {Q.~Y.}\ \bibnamefont
  {He}}, \bibinfo {author} {\bibfnamefont {S.-G.}\ \bibnamefont {Peng}},
  \bibinfo {author} {\bibfnamefont {P.~D.}\ \bibnamefont {Drummond}}, \ and\
  \bibinfo {author} {\bibfnamefont {M.~D.}\ \bibnamefont {Reid}},\ }\href
  {\doibase 10.1103/PhysRevA.84.022107} {\bibfield  {journal} {\bibinfo
  {journal} {Physical Review A}\ }\textbf {\bibinfo {volume} {84}},\ \bibinfo
  {pages} {022107} (\bibinfo {year} {2011})}\BibitemShut {NoStop}%
\bibitem [{\citenamefont {S{\o}rensen}\ and\ \citenamefont
  {M{\o}lmer}(2001)}]{Sorensen2001a}%
  \BibitemOpen
  \bibfield  {author} {\bibinfo {author} {\bibfnamefont {A.~S.}\ \bibnamefont
  {S{\o}rensen}}\ and\ \bibinfo {author} {\bibfnamefont {K.}~\bibnamefont
  {M{\o}lmer}},\ }\href {\doibase 10.1103/PhysRevLett.86.4431} {\bibfield
  {journal} {\bibinfo  {journal} {Physical Review Letters}\ }\textbf {\bibinfo
  {volume} {86}},\ \bibinfo {pages} {4431} (\bibinfo {year}
  {2001})}\BibitemShut {NoStop}%
\bibitem [{\citenamefont {{Frisk Kockum}}\ \emph {et~al.}(2019)\citenamefont
  {{Frisk Kockum}}, \citenamefont {Miranowicz}, \citenamefont {{De Liberato}},
  \citenamefont {Savasta},\ and\ \citenamefont {Nori}}]{FriskKockum2019}%
  \BibitemOpen
  \bibfield  {author} {\bibinfo {author} {\bibfnamefont {A.}~\bibnamefont
  {{Frisk Kockum}}}, \bibinfo {author} {\bibfnamefont {A.}~\bibnamefont
  {Miranowicz}}, \bibinfo {author} {\bibfnamefont {S.}~\bibnamefont {{De
  Liberato}}}, \bibinfo {author} {\bibfnamefont {S.}~\bibnamefont {Savasta}}, \
  and\ \bibinfo {author} {\bibfnamefont {F.}~\bibnamefont {Nori}},\ }\href
  {\doibase 10.1038/s42254-018-0006-2} {\bibfield  {journal} {\bibinfo
  {journal} {Nature Reviews Physics}\ }\textbf {\bibinfo {volume} {1}},\
  \bibinfo {pages} {19} (\bibinfo {year} {2019})}\BibitemShut {NoStop}%
\bibitem [{\citenamefont {H{\"{a}}u{\ss}ler}\ \emph {et~al.}(2017)\citenamefont
  {H{\"{a}}u{\ss}ler}, \citenamefont {Thiering}, \citenamefont {Dietrich},
  \citenamefont {Waasem}, \citenamefont {Teraji}, \citenamefont {Isoya},
  \citenamefont {Iwasaki}, \citenamefont {Hatano}, \citenamefont {Jelezko},
  \citenamefont {Gali},\ and\ \citenamefont {Kubanek}}]{Haussler2017}%
  \BibitemOpen
  \bibfield  {author} {\bibinfo {author} {\bibfnamefont {S.}~\bibnamefont
  {H{\"{a}}u{\ss}ler}}, \bibinfo {author} {\bibfnamefont {G.}~\bibnamefont
  {Thiering}}, \bibinfo {author} {\bibfnamefont {A.}~\bibnamefont {Dietrich}},
  \bibinfo {author} {\bibfnamefont {N.}~\bibnamefont {Waasem}}, \bibinfo
  {author} {\bibfnamefont {T.}~\bibnamefont {Teraji}}, \bibinfo {author}
  {\bibfnamefont {J.}~\bibnamefont {Isoya}}, \bibinfo {author} {\bibfnamefont
  {T.}~\bibnamefont {Iwasaki}}, \bibinfo {author} {\bibfnamefont
  {M.}~\bibnamefont {Hatano}}, \bibinfo {author} {\bibfnamefont
  {F.}~\bibnamefont {Jelezko}}, \bibinfo {author} {\bibfnamefont
  {A.}~\bibnamefont {Gali}}, \ and\ \bibinfo {author} {\bibfnamefont
  {A.}~\bibnamefont {Kubanek}},\ }\href {\doibase 10.1088/1367-2630/aa73e5}
  {\bibfield  {journal} {\bibinfo  {journal} {New Journal of Physics}\ }\textbf
  {\bibinfo {volume} {19}},\ \bibinfo {pages} {063036} (\bibinfo {year}
  {2017})}\BibitemShut {NoStop}%
\bibitem [{\citenamefont {Iwasaki}\ \emph {et~al.}(2017)\citenamefont
  {Iwasaki}, \citenamefont {Miyamoto}, \citenamefont {Taniguchi}, \citenamefont
  {Siyushev}, \citenamefont {Metsch}, \citenamefont {Jelezko},\ and\
  \citenamefont {Hatano}}]{Iwasaki2017}%
  \BibitemOpen
  \bibfield  {author} {\bibinfo {author} {\bibfnamefont {T.}~\bibnamefont
  {Iwasaki}}, \bibinfo {author} {\bibfnamefont {Y.}~\bibnamefont {Miyamoto}},
  \bibinfo {author} {\bibfnamefont {T.}~\bibnamefont {Taniguchi}}, \bibinfo
  {author} {\bibfnamefont {P.}~\bibnamefont {Siyushev}}, \bibinfo {author}
  {\bibfnamefont {M.~H.}\ \bibnamefont {Metsch}}, \bibinfo {author}
  {\bibfnamefont {F.}~\bibnamefont {Jelezko}}, \ and\ \bibinfo {author}
  {\bibfnamefont {M.}~\bibnamefont {Hatano}},\ }\href {\doibase
  10.1103/PhysRevLett.119.253601} {\bibfield  {journal} {\bibinfo  {journal}
  {Physical Review Letters}\ }\textbf {\bibinfo {volume} {119}},\ \bibinfo
  {pages} {253601} (\bibinfo {year} {2017})}\BibitemShut {NoStop}%
\bibitem [{\citenamefont {Yano}\ \emph {et~al.}(1992)\citenamefont {Yano},
  \citenamefont {Mitsunaga},\ and\ \citenamefont {Uesugi}}]{Yano1992}%
  \BibitemOpen
  \bibfield  {author} {\bibinfo {author} {\bibfnamefont {R.}~\bibnamefont
  {Yano}}, \bibinfo {author} {\bibfnamefont {M.}~\bibnamefont {Mitsunaga}}, \
  and\ \bibinfo {author} {\bibfnamefont {N.}~\bibnamefont {Uesugi}},\ }\href
  {\doibase 10.1364/JOSAB.9.000992} {\bibfield  {journal} {\bibinfo  {journal}
  {Journal of the Optical Society of America B}\ }\textbf {\bibinfo {volume}
  {9}},\ \bibinfo {pages} {992} (\bibinfo {year} {1992})}\BibitemShut {NoStop}%
\bibitem [{\citenamefont {Kolesov}\ \emph {et~al.}(2012)\citenamefont
  {Kolesov}, \citenamefont {Xia}, \citenamefont {Reuter}, \citenamefont
  {St{\"{o}}hr}, \citenamefont {Zappe}, \citenamefont {Meijer}, \citenamefont
  {Hemmer},\ and\ \citenamefont {Wrachtrup}}]{Kolesov2012}%
  \BibitemOpen
  \bibfield  {author} {\bibinfo {author} {\bibfnamefont {R.}~\bibnamefont
  {Kolesov}}, \bibinfo {author} {\bibfnamefont {K.}~\bibnamefont {Xia}},
  \bibinfo {author} {\bibfnamefont {R.}~\bibnamefont {Reuter}}, \bibinfo
  {author} {\bibfnamefont {R.}~\bibnamefont {St{\"{o}}hr}}, \bibinfo {author}
  {\bibfnamefont {A.}~\bibnamefont {Zappe}}, \bibinfo {author} {\bibfnamefont
  {J.}~\bibnamefont {Meijer}}, \bibinfo {author} {\bibfnamefont
  {P.}~\bibnamefont {Hemmer}}, \ and\ \bibinfo {author} {\bibfnamefont
  {J.}~\bibnamefont {Wrachtrup}},\ }\href {\doibase 10.1038/ncomms2034}
  {\bibfield  {journal} {\bibinfo  {journal} {Nature Communications}\ }\textbf
  {\bibinfo {volume} {3}},\ \bibinfo {pages} {1029} (\bibinfo {year}
  {2012})}\BibitemShut {NoStop}%
\bibitem [{\citenamefont {Nilsson}\ \emph {et~al.}(2004)\citenamefont
  {Nilsson}, \citenamefont {Rippe}, \citenamefont {Kr{\"{o}}ll}, \citenamefont
  {Klieber},\ and\ \citenamefont {Suter}}]{Nilsson2004}%
  \BibitemOpen
  \bibfield  {author} {\bibinfo {author} {\bibfnamefont {M.}~\bibnamefont
  {Nilsson}}, \bibinfo {author} {\bibfnamefont {L.}~\bibnamefont {Rippe}},
  \bibinfo {author} {\bibfnamefont {S.}~\bibnamefont {Kr{\"{o}}ll}}, \bibinfo
  {author} {\bibfnamefont {R.}~\bibnamefont {Klieber}}, \ and\ \bibinfo
  {author} {\bibfnamefont {D.}~\bibnamefont {Suter}},\ }\href {\doibase
  10.1103/PhysRevB.70.214116} {\bibfield  {journal} {\bibinfo  {journal}
  {Physical Review B}\ }\textbf {\bibinfo {volume} {70}},\ \bibinfo {pages}
  {214116} (\bibinfo {year} {2004})}\BibitemShut {NoStop}%
\bibitem [{\citenamefont {Agladze}\ \emph {et~al.}(1991)\citenamefont
  {Agladze}, \citenamefont {Popova}, \citenamefont {Zhizhin}, \citenamefont
  {Egorov},\ and\ \citenamefont {Petrova}}]{Agladze1991}%
  \BibitemOpen
  \bibfield  {author} {\bibinfo {author} {\bibfnamefont {N.}~\bibnamefont
  {Agladze}}, \bibinfo {author} {\bibfnamefont {M.}~\bibnamefont {Popova}},
  \bibinfo {author} {\bibfnamefont {G.}~\bibnamefont {Zhizhin}}, \bibinfo
  {author} {\bibfnamefont {V.}~\bibnamefont {Egorov}}, \ and\ \bibinfo {author}
  {\bibfnamefont {M.}~\bibnamefont {Petrova}},\ }\href {\doibase
  10.1103/PhysRevLett.66.477} {\bibfield  {journal} {\bibinfo  {journal}
  {Physical Review Letters}\ }\textbf {\bibinfo {volume} {66}},\ \bibinfo
  {pages} {477} (\bibinfo {year} {1991})}\BibitemShut {NoStop}%
\bibitem [{\citenamefont {Shakhmuratov}\ \emph {et~al.}(2005)\citenamefont
  {Shakhmuratov}, \citenamefont {Rebane}, \citenamefont {M{\'{e}}gret},\ and\
  \citenamefont {Odeurs}}]{Shakhmuratov2005}%
  \BibitemOpen
  \bibfield  {author} {\bibinfo {author} {\bibfnamefont {R.~N.}\ \bibnamefont
  {Shakhmuratov}}, \bibinfo {author} {\bibfnamefont {A.}~\bibnamefont
  {Rebane}}, \bibinfo {author} {\bibfnamefont {P.}~\bibnamefont
  {M{\'{e}}gret}}, \ and\ \bibinfo {author} {\bibfnamefont {J.}~\bibnamefont
  {Odeurs}},\ }\href {\doibase 10.1103/PhysRevA.71.053811} {\bibfield
  {journal} {\bibinfo  {journal} {Physical Review A}\ }\textbf {\bibinfo
  {volume} {71}},\ \bibinfo {pages} {053811} (\bibinfo {year}
  {2005})}\BibitemShut {NoStop}%
\bibitem [{\citenamefont {Sabooni}\ \emph {et~al.}(2013)\citenamefont
  {Sabooni}, \citenamefont {Li}, \citenamefont {Rippe}, \citenamefont {Mohan},\
  and\ \citenamefont {Kr{\"{o}}ll}}]{Sabooni2013}%
  \BibitemOpen
  \bibfield  {author} {\bibinfo {author} {\bibfnamefont {M.}~\bibnamefont
  {Sabooni}}, \bibinfo {author} {\bibfnamefont {Q.}~\bibnamefont {Li}},
  \bibinfo {author} {\bibfnamefont {L.}~\bibnamefont {Rippe}}, \bibinfo
  {author} {\bibfnamefont {R.~K.}\ \bibnamefont {Mohan}}, \ and\ \bibinfo
  {author} {\bibfnamefont {S.}~\bibnamefont {Kr{\"{o}}ll}},\ }\href {\doibase
  10.1103/PhysRevLett.111.183602} {\bibfield  {journal} {\bibinfo  {journal}
  {Physical Review Letters}\ }\textbf {\bibinfo {volume} {111}},\ \bibinfo
  {pages} {183602} (\bibinfo {year} {2013})}\BibitemShut {NoStop}%
\bibitem [{\citenamefont {Utzat}\ \emph {et~al.}(2019)\citenamefont {Utzat},
  \citenamefont {Sun}, \citenamefont {Kaplan}, \citenamefont {Krieg},
  \citenamefont {Ginterseder}, \citenamefont {Spokoyny}, \citenamefont {Klein},
  \citenamefont {Shulenberger}, \citenamefont {Perkinson}, \citenamefont
  {Kovalenko},\ and\ \citenamefont {Bawendi}}]{Utzat2019}%
  \BibitemOpen
  \bibfield  {author} {\bibinfo {author} {\bibfnamefont {H.}~\bibnamefont
  {Utzat}}, \bibinfo {author} {\bibfnamefont {W.}~\bibnamefont {Sun}}, \bibinfo
  {author} {\bibfnamefont {A.~E.~K.}\ \bibnamefont {Kaplan}}, \bibinfo {author}
  {\bibfnamefont {F.}~\bibnamefont {Krieg}}, \bibinfo {author} {\bibfnamefont
  {M.}~\bibnamefont {Ginterseder}}, \bibinfo {author} {\bibfnamefont
  {B.}~\bibnamefont {Spokoyny}}, \bibinfo {author} {\bibfnamefont {N.~D.}\
  \bibnamefont {Klein}}, \bibinfo {author} {\bibfnamefont {K.~E.}\ \bibnamefont
  {Shulenberger}}, \bibinfo {author} {\bibfnamefont {C.~F.}\ \bibnamefont
  {Perkinson}}, \bibinfo {author} {\bibfnamefont {M.~V.}\ \bibnamefont
  {Kovalenko}}, \ and\ \bibinfo {author} {\bibfnamefont {M.~G.}\ \bibnamefont
  {Bawendi}},\ }\href {\doibase 10.1126/science.aau7392} {\bibfield  {journal}
  {\bibinfo  {journal} {Science}\ }\textbf {\bibinfo {volume} {363}},\ \bibinfo
  {pages} {1068} (\bibinfo {year} {2019})}\BibitemShut {NoStop}%
\bibitem [{\citenamefont {Quang}\ and\ \citenamefont
  {Freedhoff}(1994)}]{Quang1994}%
  \BibitemOpen
  \bibfield  {author} {\bibinfo {author} {\bibfnamefont {T.}~\bibnamefont
  {Quang}}\ and\ \bibinfo {author} {\bibfnamefont {H.}~\bibnamefont
  {Freedhoff}},\ }\href {\doibase 10.1016/0030-4018(94)90366-2} {\bibfield
  {journal} {\bibinfo  {journal} {Optics Communications}\ }\textbf {\bibinfo
  {volume} {107}},\ \bibinfo {pages} {480} (\bibinfo {year}
  {1994})}\BibitemShut {NoStop}%
\end{thebibliography}%
\end{document}